\documentstyle[11pt,epsf]{article}

\newcommand{\newsection}{ \setcounter{equation}{0} \section}

\newcommand{\beq}{\begin{equation}} \newcommand{\eeq}{\end{equation}}
\newcommand{\bea}{\begin{eqnarray}} \newcommand{\eea}{\end{eqnarray}}

  \newcommand
{\Romannumeral}[1]{\uppercase\expandafter{\romannumeral#1}}

\newcommand{\be}{\begin{enumerate}} \newcommand{\ee}{\end{enumerate}}
\newcommand{\bi}{\begin{itemize}} \newcommand{\ei}{\end{itemize}}
\newcommand{\ba}{\begin{array}} \newcommand{\ea}{\end{array}}
\newcommand{\bc}{\begin{center}} \newcommand{\ec}{\end{center}}
\newcommand{\bt}{\begin{tabular}} \newcommand{\et}{\end{tabular}}

\def\lsim{\mathrel{\rlap{\lower4pt\hbox{\hskip1pt$\sim$}}
    \raise1pt\hbox{$<$}}}           
\def\gsim{\mathrel{\rlap{\lower4pt\hbox{\hskip1pt$\sim$}}
    \raise1pt\hbox{$>$}}}           

\newcommand{\half}{\textstyle {1\over2} \displaystyle}    
\newcommand{\Dslash}{{\hbox{D}\kern-0.6em\raise0.15ex\hbox{/}}} 

\renewcommand{\d}{\delta} 
 
\renewcommand{\l}{\lambda}

\begin{document}

\setlength{\oddsidemargin}{0cm} \setlength{\baselineskip}{7mm}

\input epsf




\begin{normalsize}\begin{flushright}
    UCI-04-43 \\
    May 2004 \\
\end{flushright}\end{normalsize}

\begin{center}
  
\vspace{30pt}
  
{\Large \bf Exact Bianchi Identity in Regge Gravity}

\vspace{40pt}
  
{\sl Herbert W. Hamber}
$^{}$\footnote{e-mail address : hhamber@uci.edu}
and {\sl Geoff Kagel}
$^{}$\footnote{e-mail address : gkagel@uci.edu} \\

Department of Physics and Astronomy \\
University of California \\
Irvine, CA 92697-4575, USA \\

\end{center}

\vspace{20pt}

\begin{center} {\bf ABSTRACT } \end{center}
\vspace{12pt}
\noindent

In the continuum the Bianchi identity implies a relationship
between different components of the curvature tensor, thus ensuring
the internal consistency of the gravitational field equations.
In this paper the exact form for the Bianchi identity in Regge's 
discrete formulation of gravity is derived, by considering
appropriate products of rotation matrices constructed 
around null-homotopic paths.
The discrete Bianchi identity implies an algebraic relationship between deficit angles
belonging to neighboring hinges.
As in the continuum, the derived identity is valid for
arbitrarily curved manifolds without a restriction to the weak field
small curvature limit, but is in general not linear in the curvatures.

\vspace{24pt}

\begin{center}{\it (Submitted to Class. and Quant. Gravity) }\end{center}

\vfill

\pagestyle{empty}

\newpage

\pagestyle{plain}

\vskip 10pt
\newsection{Introduction}
\hspace*{\parindent}

In this paper we investigate the form of the Bianchi identities
in Regge's \cite{regge} lattice formulation of gravity
\cite{whe64,rw81,rw82,cms,cms1,hw84,ha1,ha2,lesh}. 
The Bianchi identities play an important role in the continuum
formulation of gravity, both classical and quantum-mechanical,
giving rise to a differential relationship between different components
of the curvature tensor. It is known that these simply follow from
the definition of the Riemann tensor in terms of the affine connection and
the metric components, and help ensure the consistency of the gravitational
field equations in the presence of matter. At the same time
they can be regarded as a direct consequence of the local gauge
(diffeomorphism) invariance of the gravitational action, since they
can be derived by invoking the invariance of the action under infinitesimal 
gauge transformations (see for example \cite{fey,weibook,mtw,wald}).

In this paper we will show that the lattice formulation
of gravity has an equivalent form of the Bianchi identities,
which are both exact, in the sense that they are
valid for arbitrarily curved lattices, and reduce to their continuum
counterparts in the weak field limit.
We will derive the exact lattice Bianchi identities in three (Section 4)
and four dimensions (Section 5)
explicitly by considering the product of rotation matrices along paths
which are topologically trivial (i.e. reducible to a point).
By expressing the rotation matrices about each hinge in terms
of the local representatives of the curvatures, namely the deficit 
angles, we will obtain an algebraic relationship between
deficit angles, area and volumes pertaining to neighboring
simplices.

For lattices which are close to flat, it will be shown
that the derived set of identities is analogous to the Bianchi
identities in the continuum, once the edge lengths are
identified with appropriate components of the metric
in the continuum (Sections 9 and 10).  
We will therefore extend and complete previous results
on the lattice Bianchi identities, which so far have been
restricted to the weak field limit \cite{regge,rw81,rw82,miller,morse}.
The results presented in this paper should therefore
be relevant for both classical (see for example
\cite{class,sork,class1,wt} an references therein) and quantum
(see for example \cite{rw85,monte,rw86,pir,jn,ban,rw91,hw3d,imm,hmw,gauge}
and references therein)
discrete gravity.
\footnote{Recent reviews of Regge gravity can be found in 
\cite{lesh,leerev,cargese,rw92,froh,regge00}, while a comprehensive 
collection of up-to-date references is assembled in \cite{refs}.
Some further mathematical aspects of piecewise linear spaces, with
some relevance to lattice gravity, are discussed in the above cited references
\cite{cms,cms1}, as well as
in \cite{cox,cox1,alex1,alex2,thurst,kirby,warner,chee,cast} and references
therein, and more 
recently in \cite{arch,bar2,bar3,arci,carf,khat} and references therein.}

It is well known that in the continuum the Bianchi identity for the curvature tensor
ensures the consistency of the Einstein field equations.
For the Riemann curvature tensor the un-contracted Bianchi
identities read
\beq
R^{\mu}_{\;\; \nu \alpha \beta  ; \, \gamma} +
R^{\mu}_{\;\; \nu \gamma \alpha ; \, \beta } +
R^{\mu}_{\;\; \nu \beta  \gamma ; \, \alpha} \; = \; 0 \; ,
\label{eq:bianchi}
\eeq
or more concisely
\beq
R^{\mu}_{\;\; \nu [ \alpha \beta  ; \, \gamma ]} \; = \; 0 \; , 
\eeq
where $[\cdots]$ denotes symmetrization.
These identities are easily derived by inserting into the above expression
the explicit definition for the curvature tensor in terms of the
metric $g_{\mu\nu}$,
\beq
R_{\lambda \mu \nu \kappa} \; = \; {1 \over 2} \; {\partial^2 g_{\lambda\nu}
\over \partial x^{\kappa} \partial x^{\mu}} + \cdots \;\; .
\eeq
It is then easy to see that in $d$ dimensions there are 
\beq
{d (d-1) \over 2} \; \left ( d \atop 3 \right ) \; = \;
{d^2 (d-1)(d-2) \over 12}
\eeq
un-contracted Bianchi identities, and thus 3 identities in $d=3$
and 24 identities in $d=4$.
In their contracted form, the Bianchi identities imply for the Ricci tensor
\beq
R_{\nu \alpha ; \, \gamma} -
R_{\nu \gamma ; \, \alpha} +
R^{\mu}_{\;\; \nu \gamma \alpha ; \, \mu} \; = \; 0 \; ,
\eeq
and for the scalar curvature
\beq
R_{; \, \gamma} - 2 R^{\mu}_{\;\; \gamma ; \, \mu} \; = \; 0 \; .
\eeq
These relations in turn give the contracted Bianchi identity
\beq
\left [ \, R^{\mu}_{\;\; \nu} - \half \, \delta^{\mu}_{\;\; \nu} \, R
\, \right ]_{; \, \mu} \; = \; 0 \; ,
\label{eq:bianchi_cont}
\eeq
which always corresponds to $d$ equations in $d$ dimensions.
We note here that a simple physical interpretation for the Bianchi identity can
be given in terms of a divergence of suitably defined stresses \cite{fey}.
Thus for example in three dimensions one has
\beq
P^{\mu}_{\;\; \nu i; \, i}  \; = \; 0 \;\;\;\; {\rm with} \;\;\;\;
P^{\mu}_{\;\; \nu i}  \; \equiv \; \epsilon_{ijk} R^{\mu}_{\;\; \nu j k} \;\; .
\label{eq:P}
\eeq
It is also well known that the Bianchi identities are required for ensuring the 
consistency of the gravitational field equations.
Consider the classical field equations with a cosmological constant term,
\beq
R_{\mu\nu} - \half g_{\mu\nu} R + \Lambda g_{\mu\nu}
\; = \; 8 \pi G \, T_{\mu\nu} ,
\label{eq:field_eq}
\eeq
with $ \Lambda = 8 \pi G \lambda $ the cosmological constant.
Applying a covariant derivative on both sides one has
\beq
\left [ R^{\mu}_{\;\; \nu} - \half \, \delta^{\mu}_{\;\; \nu} \, R
\right ]_{; \, \mu} \; = \; 8 \pi G \, T^{\mu}_{\;\; \nu ; \, \mu} ,
\eeq
which for a covariantly conserved energy-momentum tensor
\beq
T^{\mu}_{\;\; \nu ; \, \mu} \; = \; 0 \; ,
\eeq
is only consistent if the contracted Bianchi identity 
of Eq.~(\ref{eq:bianchi_cont}) is identically satisfied.

In $d$ dimensions one has $d$ contracted Bianchi identities.
Since there are in general $d(d+1)/2$ equations of motion, as well as
$d$ harmonic gauge fixing conditions, one has for the number of
independent gravitational degrees of freedom in $d$ dimensions
\beq
{d (d+1) \over 2} \; - \; d \; - \; d \; = \; {d(d-3) \over 2}
\eeq
which indeed reproduces correctly in four dimensions the two independent helicity states
appropriate for a massless spin two particle.

There exists also a close relationship between the Bianchi identity
and the gauge invariance of the gravitational action.
We shall take note here of the fact that the Bianchi identity can
be derived from the requirement that the gravitational action
\beq
I_G [g] \; = \; - {1 \over 16 \pi G} \int d^4 x \,
{\textstyle \sqrt{g(x)} \displaystyle} \, R(x) \;\; ,
\label{eq:einst_ac}
\eeq
being a scalar, should be invariant under infinitesimal local 
gauge transformations at a space-time point $x$,
\beq
\delta g_{\mu\nu} (x) \, = \, 
- g_{\mu\lambda} (x) \, \partial_\nu \chi^\lambda (x)
- g_{\lambda\nu} (x) \, \partial_\mu \chi^\lambda (x)
- \partial_{\lambda} g_{\mu\nu} (x) \, \chi^\lambda (x) \;\; .
\label{eq:var_g}
\eeq
After substituting the above expression for a gauge deformation
into the variation of the action given in Eq.~(\ref{eq:einst_ac}),
\beq
\delta \, I_G [g] \; = \; {1 \over 16 \pi G} \int d^4 x \,
{\textstyle \sqrt{g(x)} \displaystyle} \, \left [ R^{\mu\nu} - \half 
g^{\mu\nu} R \right ] \, \delta g_{\mu\nu} \;\; ,
\eeq
one obtains again, after integrating by parts, the contracted Bianchi
identity of Eq.~(\ref{eq:bianchi_cont}).

Let us now turn to the lattice theory.
In a $d$-dimensional piecewise linear space-time the expression
analogous to the Einstein-Hilbert action was given by Regge \cite{regge} as
\beq
I_R \; = \; \sum_ {{\bf hinges \; h}} A_h^{(d-2)} \; \delta_h \;\; ,
\label{eq:regge_ac}
\eeq
where $A_h^{(d-2)} $ is the volume of the hinge and $\delta_h$ is
the deficit angle there.
The above lattice action is supposed to be equivalent to
the continuum expression
\beq
I_E \; = \; {1 \over 2} \int d^d x \sqrt g \; R  \;\; ,
\eeq
and indeed it has been
shown \cite{rw81,cms,lee} that $ I_R $ tends to the continuum expression
as the simplicial block size (or some suitable average edge length) tends to zero
in the appropriate way.
The above Regge form for the lattice action can be naturally extended to include
cosmological and curvature squared terms \cite{hw84,lesh}
\beq
I (l^2) \; = \; \sum_h \, \Bigl [ \lambda \, V_h - k \, A_h \delta_h 
+ a \, \delta_h^2 A_h^2 / V_h + \cdots \Bigr ] \;\; ,
\label{eq:pure} 
\eeq
with $k^{-1} = 8 \pi G $.
In the limit of small fluctuations around a smooth background, $I(l^2)$
corresponds to the continuum action
\beq
I [g] \; = \; 
\int d^4 x \, \sqrt g \, \Bigl [ \lambda - {k \over 2} \, R
+ {a \over 4} \, R_{\mu\nu\rho\sigma} R^{\mu\nu\rho\sigma} + \cdots
\Bigr ] \;\; .
\eeq
In the following we will focus on the Regge term (proportional
to $k$) only.

Variations of $ I_R $ in Eq.~(\ref{eq:regge_ac})
with respect to the edge lengths then
give the simplicial analogs of Einstein's equations, whose derivation is
significantly simplified by the fact that the variation of the deficit angle
is known to be zero in any dimensions,
\beq
\delta I_R \; = \; \sum_ h \delta ( A_h^{(d-2)} ) \; \delta_h \;\; ,
\eeq
as happens in the continuum as well (where one finds that the
variation of the curvature reduces to a total derivative).
In three dimensions the above action gives for the equation of motion
$ \delta_h = 0 $ for every hinge in the lattice, 
whereas in four dimensions variation with respect to $l_p $ yields \cite{regge}
\beq
{1 \over 2} \; l_p 
\sum_ {h \supset l_p} \delta_h \cot \theta_{ph} \; = \; 0
\label{eq:regge_eq}
\eeq
where the sum is over hinges (triangles in four dimensions)
labeled by $h$ meeting on the
common edge $p$, and $\theta_{ph} $ is the angle in the hinge $h$
opposite to the edge $p$.
The above equation is the lattice analog of the field equations of Eq.~(\ref{eq:field_eq}),
for pure gravity and vanishing cosmological constant.

Numerical solutions to the lattice equations of motion
can in general be found by appropriately adjusting the edge lengths
according to Eq.~(\ref{eq:regge_eq}).
Since the resulting equations are non-linear in the edge lengths,
slight complications can arise such as the existence
of multiple solutions, although
for sufficiently weak fields one would expect the same
level of degeneracies as in the continuum \cite{rw81}.
Several authors have discussed the application of the Regge equations to
strong field problems in classical general relativity, and some samples can be
found in \cite{class,sork,class1}.
The relevance of the Bianchi identities to a numerical solution
to the lattice field equations - using for example a $3+1$ time
evolution scheme - resides in the fact that they are in principle
a powerful tool to check the overall accuracy and consistency of the numerical solutions.

The Bianchi identities also play an important role in the quantum
formulation.
In a quantum-mechanical theory of gravity the starting point is 
a suitable definition of the discrete Feynman path integral
\cite{ha1,hw84,lesh}.
In the simplicial lattice approach one starts from the discretized
Euclidean Feynman path integral for pure gravity,
with the squared edge lengths taken as fundamental variables,
\beq
Z_L \; = \; \int_0^\infty \; \prod_s \; \left ( V_d (s) \right )^{\sigma} \;
\prod_{ij} \, dl_{ij}^2 \; \Theta [l_{ij}^2]  \; 
\exp \left \{
- \sum_h \, \Bigl ( \lambda \, V_h - k \, \delta_h A_h 
+ a \, \delta_h^2 A_h^2 / V_h  + \cdots \Bigr ) \right \}  \;\; .
\label{eq:zlatt} 
\eeq
The above regularized lattice expression should be compared to
the continuum Euclidean path integral for pure gravity
\beq
Z_C \; = \; \int \prod_x \;
\left ( {\textstyle \sqrt{g(x)} \displaystyle} \right )^{\sigma}
\; \prod_{\mu \ge \nu} \, d g_{\mu \nu} (x) \;
\exp \left \{
- \int d^4 x \, \sqrt g \, \Bigl ( \lambda - {k \over 2} \, R
+ {a \over 4} \, R_{\mu\nu\rho\sigma} R^{\mu\nu\rho\sigma}
+ \cdots \Bigr ) \right \}  \;\; .
\label{eq:zcont}
\eeq
In the discrete case the integration over metrics is replaced by
integrals over the elementary lattice degrees of freedom,
the squared edge lengths.
The discrete gravitational measure in $Z_L$
can be considered as the lattice analog of the DeWitt
continuum functional measure \cite{cms,cms1,ha1,ha2,leerev,gauge}.
A cosmological constant term is needed for convergence of the path
integral, while the curvature squared term allows one to control the
fluctuations in the curvature \cite{hw84,lesh}.
In the end one is mostly interested in the limit $a \rightarrow 0$,
where the theory, in the
absence of matter and after a suitable rescaling of the metric, only 
depends on one bare parameter, the dimensionless coupling $k^2 / \lambda $.

In the quantum theory the Bianchi identities of Eq.~(\ref{eq:bianchi_cont})
are still satisfied as operator equations, and ensure the consistency
of the quantum equations of motions. 
In ordinary lattice nonabelian gauge theories an attempt
has been made to entirely replace the functional
integration over the gauge fields by an integration over field
strengths, but now subject to the Bianchi identity constraint \cite{lgt1}.
In the case of gravitation such an approach is more difficult, since
the analog of the gauge field is represented by the affine connection, and not
the by the curvature tensor.

\vskip 30pt
\newsection{Lattice Parallel Transport}
\hspace*{\parindent}

To construct the lattice Bianchi identities we will follow a strategy
similar to the one used in the derivation of the exact lattice Bianchi
identities in non-abelian lattice gauge theories.
There the Bianchi
identities can be obtained by considering the path-ordered
product of $SU(n)$ gauge group rotation matrices, taken along a suitable closed path
encircling a cube.
The path has to be chosen topologically trivial, in the sense
that it can be shrunk to a point without entangling any
plaquettes \cite{lgt1,lgt2}.

Let us therefore first review the notion of parallel transport
of a test vector around a small loop embedded in the lattice.
Consider a closed path $\Gamma$ encircling a hinge $h$ and passing through
each of the simplices that meet at that hinge.
In particular one may take $ \Gamma$ to be the boundary of the polyhedral
dual area surrounding the hinge.
For each neighboring pair of simplices $j,j+1$, one can write down a
Lorentz transformation $L_\mu^{\; \nu}$, which describes how a given
vector $ \phi_\mu $ transforms between the local coordinate systems in
these two simplices,
\beq
{\phi '}_\mu \; = \; \Bigl [ L (j,j+1) \Bigr ]_\mu^{\;\; \nu} \phi_\nu \;\; .
\label{eq:latt_rot1}
\eeq
Now in general it is possible to choose coordinates so that
$ L_\mu^{\; \nu} $ is the identity matrix for one pair of simplices,
but then it will not be unity for other pairs.
The above Lorentz transformation is directly related to the continuum
path-ordered
($P$) exponential of the integral of the affine connection
$ ( \Gamma_\lambda )_\mu^{\nu} =
\Gamma_{\mu \lambda}^\nu $ by
\beq
L_\mu^{\;\; \nu} \; = \; \Bigl [ P e^{\int_
{{\bf path \atop between \; simplices}}
\Gamma_\lambda d x^\lambda} \Bigr ]_\mu^{\;\; \nu}  \;\; .
\label{eq:cont_rot1}
\eeq
The connection here is intended to only have support on the common
interface between the two simplices.

Next we will consider the product of rotation matrices along a {\it closed}
loop $\Gamma$.
The path can entangle several hinges, or just one, in which case
it will be called a closed elementary loop.
On the lattice the effect of parallel transport around
a closed elementary loop $\Gamma$ is obtained from the matrix \cite{hw84}
\beq
\Bigl [ \prod_j L (j,j+1) \Bigr ]_{\mu \nu} =
\Bigl [ e^{\delta_h U_{..}^{(h)}} \Bigr ]_{\mu \nu}  \;\; ,
\label{eq:fullrot1}
\eeq
where $U_{\mu \nu}^{(h)} $ is a bivector orthogonal to the hinge $h$,
defined in four dimensions by
\beq
U_{\mu \nu}^{(h)} \; = \; {1 \over 2 A_h} 
\; \epsilon_{\mu \nu \rho \sigma} \; l_{(a)}^\rho l_{(b)}^\sigma  \;\; ,
\label{eq:bivector}
\eeq
with $ l_{(a)}^\rho $ and $ l_{(b)}^\rho $ two vectors forming
two sides of the hinge $h$.
We note that in general the validity of the lattice parallel
transport formula given above is not restricted to small deficit angles.
For a closed path $\Gamma$, the total change in a
vector $ \phi_\mu $ which undergoes parallel transport around
the path is given by
\beq
{\phi '}_\mu \; = \; \phi_\mu + \delta \phi_\mu \; = \; \Bigl [ \prod_
{{\bf pairs \; of \atop simplices \; on \; \Gamma}}
L (j,j+1) \Bigr ]_\mu^{\;\; \nu} \phi_\nu
\label{eq:fullrot2}
\eeq
For smooth enough manifolds, the product of Lorentz transformations
around a closed elementary loop $\Gamma$ can be deduced from
the components of the Riemann tensor,
\beq
\Bigl [ \prod_{{\bf pairs \; of \atop simplices \; on \; \Gamma}}
L (j,j+1) \Bigr ] 
_ \mu^{\;\; \nu} \approx \Bigl [ e^{R_{. \rho \sigma}^{.}
\Sigma^{\rho \sigma}} \Bigr ]_\mu^{\;\; \nu} \;\; ,
\label{eq:fullrot3}
\eeq
where $( R^{.}_{\; . \rho \sigma} )_\mu^{\nu} =
R_{\; \mu \rho \sigma}^{\nu}$
is the curvature tensor and 
$ \Sigma^{\rho \sigma} $ is a bivector in the plane of $\Gamma$,
with magnitude equal to $1/ 2 $ times the area of the loop $\Gamma$.
(For a parallelogram
with edges $a^\rho$ and $ b^\rho $, $\Sigma^{\rho \sigma} =
{1 \over 2} ( a^\sigma b^\rho - a^\rho b^\sigma $)).
The above result then reproduces to lowest order the parallel transport formula
\beq
\delta \phi_\mu \; = \; R_{\; \mu \rho \sigma}^{\nu} 
\Sigma^{\rho \sigma} \phi_\nu  \;\; .
\eeq
Comparison of Eq.~(\ref{eq:fullrot1}) and Eq.~(\ref{eq:fullrot3})
means that for {\it one} hinge one may make the identification
\beq
R_{\mu \nu \rho \sigma} \Sigma^{\rho \sigma}
\;\; \rightarrow \;\; \delta_h U_{\mu \nu}^{(h)}  \;\; .
\eeq
It is important to notice that this relation does not give complete
information about the Riemann tensor, but only about its projection in the
plane of the loop $\Gamma$, orthogonal to the given hinge.
Indeed the deficit angle divided by the area of the loop can be taken as
a definition of the {\it local sectional curvature} $K_h$  \cite{lesh}
\beq
{\delta_h \over A_{\Gamma_h}}
= K_h \; = \; {
R_{\mu \nu \rho \sigma} \;
e_a^\mu e_b^\nu e_a^\rho e_b^\sigma
\over 
( g_{\mu \rho} g_{\nu \sigma} -
g_{\mu \sigma} g_{\nu \rho} ) 
e_a^\mu e_b^\nu e_a^\rho e_b^\sigma} \;\; ,
\eeq
which represents the projection of the Riemann curvature in the direction
of the bivector $ {\bf e_a} \wedge {\bf e_b} $.

The lattice Bianchi identities
are derived by considering closed paths that can be
shrunk to a point without entangling any hinge.
The product of rotation matrices associated with the path then has to
give the identity matrix \cite{regge,rw81}.
Thus, for example, the ordered product of rotation matrices associated with the
triangles meeting on a given edge has to give one, since a path can be
constructed which sequentially encircles all the triangles and is
topologically trivial
\beq
\prod_{{\bf hinges \; h \atop meeting \; on \; edge \; p}} 
\Bigl [ e^{\delta_h U_{..}^{(h)}} \Bigr ]_{\mu \nu} \; = \; 1  \;\; .
\label{eq:bianchi_latt}
\eeq
Other identities might be derived by considering paths that encircle hinges
meeting on one point.

\vskip 30pt
\newsection{Geometric Setup}
\hspace*{\parindent}

The discrete analogue of the Bianchi identity will be derived by considering
a product of rotation matrices along a homotypically trivial path.
This section discusses the general geometric setups needed to define 
correctly the product of rotation matrices entering the exact lattice Bianchi
identities derived later in the paper (Sections 4 and 5). 
For the 3-d case, consider a tetrahedron 
with a point in its interior.  Connect the vertices of the 
tetrahedron to the point in the center.  We now have formed 4 
tetrahedra from the original tetrahedron.  In three dimensions, 
hinges are edges, so here we have enclosed four hinges:  one connecting each 
vertex of the original tetrahedra to the interior point.

\vskip 130pt

\begin{center}
\leavevmode
\epsfysize=7cm
\epsfbox{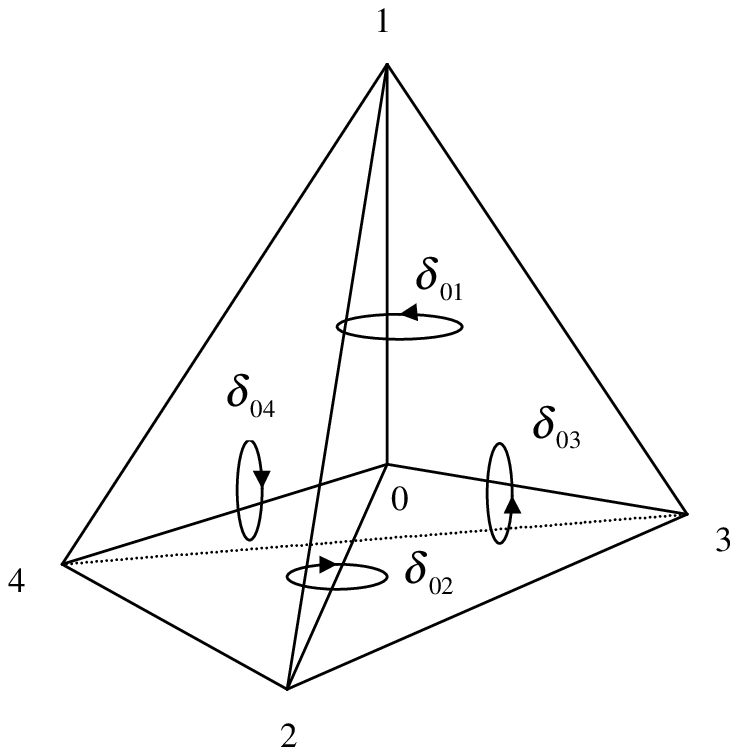}
\end{center}
\noindent
{\small{\it Fig.\ 1.
In three dimensions four tetrahedra meet on a point, labeled here by $0$.
Deficit angles $\delta_{01}$, $\delta_{02}$, $\delta_{03}$ and $\delta_{04}$
are associated with edges $0-1$, $0-2$, $0-3$ and $0-4$ respectively.
The Bianchi identities are obtained by taking an ordered product of
rotation matrices along a path which encircles all four hinges (edges here)
and is topologically trivial, in the sense that it can be shrunk to a point.
\medskip}}

\vskip 20pt

Referring to Fig. 1, for the moment only considering flat space, 
call the interior point 0 (zero) and place it at the 
origin.  Let us take our coordinate system so that vertex 1 
lies on the positive z axis ($z1>0$).  Let us take vertex 2 to lie 
in the x-z plane with $z2<0$ and $x2>0$.  Let us take vertex 3 to have 
$z3<0$, $x3<0$ and $y3>0$.  Finally, let us take vertex 4 to have 
$z4<0$, $x4<0$ and $y4<0$.  So, to summarize:

\begin{equation}
\begin{array}{ccccc}
{\rm Vertex}: & 1 & 2 & 3 & 4 \\
  z:    & + & - & - & - \\
  x:    & 0 & + & - & - \\
  y:    & 0 & 0 & + & - 
\end{array}
\label{eq:vertsigns}
\end{equation} 
Now, the following argument will apply for 
all cases where we have the center point completely surrounded; the 
aforementioned restrictions in Eq.~(\ref{eq:vertsigns}) are mentioned 
only to give the reader a nice picture of the situation.

Now, "curve the space".  This is done by changing one of the 
edge lengths.  Any 9 of the 10 edge lengths can be chosen 
arbitrarily (provided the center point is completely surrounded 
by the constructed volumes and provided that real areas and real volumes 
are still formed) 
and the space will still be flat.  
So curvature, in this setup, just amounts to adjusting one 
edge length ($l_{34}$, the edge between vertices 3 and 4, is 
the easiest one to adjust).  For an arbitrary setup, one 
adds more edges until the relative flat space locations of all vertices 
are specified, making sure not to add any edges between points whose 
relative flat space location is already determined but rather 
marking each such edge as an "unadded edge"  
(and making sure that all areas and volumes are real); 
then, all remaining ("unadded") edges are determined for 
flat space, and it is the varying of those remaining edges 
which is the source of all curvature for that setup.  This 
last comment applies to d-dimensions and general lattices, 
not just d-simplices surrounding one point, and addresses 
the question of curvature invariance under edge variation,  
i.e. curvature invariance under the edge variation 
of the added edges (with the ``unadded'' edges still unadded).

Further, let us label the four tetrahedra by a point in each of 
them along the line $const \times ( {\bf l}_1 + {\bf l}_2 + {\bf l}_3 )$ where 
${\bf l}_1$, ${\bf l}_2$ and ${\bf l}_3$ are vectors based 
at the point $0$, the interior point.

\vskip 90pt

\begin{center}
\leavevmode
\epsfysize=7cm
\epsfbox{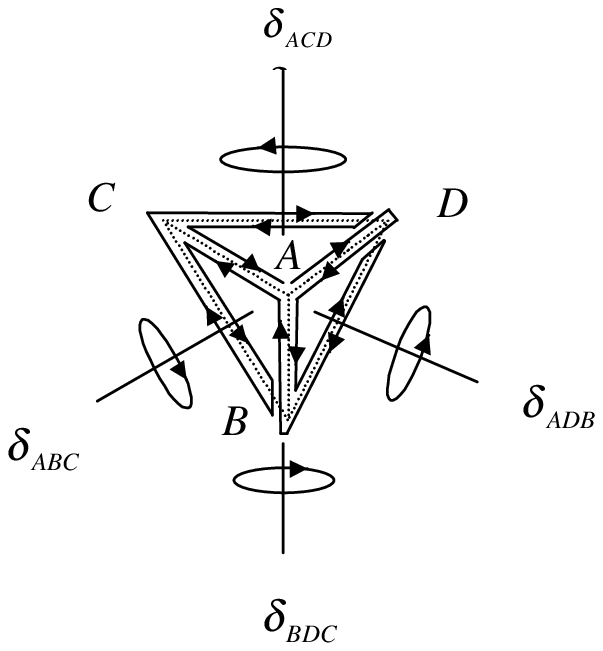}
\end{center}
\noindent
{\small{\it Fig.\ 2.
A topologically trivial path which encircles four hinges in three
dimensions, and can be shrunk to a point.
The vertices $A$, $B$, $C$ and $D$ reside in the dual lattice,
and can conveniently be placed at the centers of the tetrahedra shown
in Fig.\ 1. It is then advantageous to label the deficit angles by the vertices
in the dual lattice.
The path ordered product of rotation matrices around the shown path 
then reduces to the identity matrix.
\medskip}}

\vskip 20pt

Referring to Fig. 1, and Fig. 2 which,
with triangle BCD behind A, views Fig. 1 from underneath, 
let us call the point in the center of the tetrahedra formed 
via vertices 0123, and not 4, point D; similarly call the point 
at the center of 0234, and not 1, point A, 
that at the center of 0134, and not 2, point B and that at the center of 
0124, and not 3, point C.  
We also have tetrahedron A which is the tetrahedron containing point A, 
tetrahedron B containing point B, etc.  
These four points, point A, point B, point C and point D form a Voronoi 
tetrahedra.  Going from $A\rightarrow B\rightarrow C\rightarrow A$, for example, goes around 
hinge $04$.  In general
\begin{equation}
\begin{array}{ccccc}
\mbox{hinge gone around}  & \mbox{rotation} & \mbox{Voronoi vertex not in path} & \mbox{positive path} & \mbox{negative path}\\
01 & R1 & A & B\rightarrow C\rightarrow D\rightarrow B & B\rightarrow D\rightarrow C\rightarrow B \\
02 & R2 & B & A\rightarrow D\rightarrow C\rightarrow A & A\rightarrow C\rightarrow D\rightarrow A \\
03 & R3 & C & A\rightarrow B\rightarrow D\rightarrow A & A\rightarrow D\rightarrow B\rightarrow A \\
04 & R4 & D & A\rightarrow C\rightarrow B\rightarrow A & A\rightarrow B\rightarrow C\rightarrow A
\end{array}
\label{eq:hingenota}
\end{equation}
so that rotations which are clockwise when viewed from outside 
our setup, or counterclockwise when viewed from the point in the 
middle, are associated with a "-R", and rotations which are counter-clockwise 
when viewed from outside our setup, or clockwise when viewed from 
the point in the middle, are associated with a "+R", as is traditional 
for a right handed coordinate system inside the set-up.

\vskip 90pt

\begin{center}
\leavevmode
\epsfysize=7cm
\epsfbox{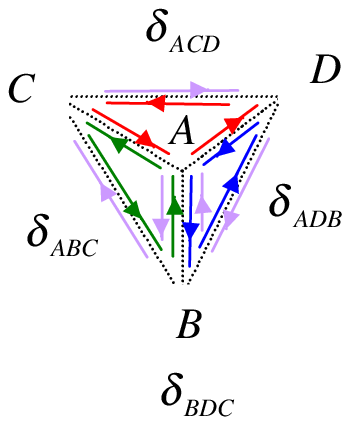}
\end{center}
\noindent
{\small{\it Fig.\ 3.
Paths associated with different hinges in three dimensions
are shown in different shades of gray.
The product of rotation matrices around a single hinge is simply 
elated to the deficit angle for that hinge.
In order to obtain a closed path, segment AB has to be transversed
more than two times.
\medskip}}
 
\vskip 20pt

Now, let us try and compose a null path around all four hinges. 
Referring to Fig. 3, we write our null path as
\begin{equation}
A\rightarrow B\rightarrow D\rightarrow A,A\rightarrow D\rightarrow C\rightarrow A,A\rightarrow C\rightarrow B\rightarrow A,A\rightarrow B,B\rightarrow C\rightarrow D\rightarrow B,B\rightarrow A
\label{eq:3dnullpath}
\end{equation}
or, noting that the path is a null path, we have 
\begin{equation}
1=R_{tot}=\Gamma_{A\rightarrow B}^{-1} \; R1_B \; \Gamma_{A\rightarrow B} \; R4 \; R2 \; R3
\label{eq:3dnullrotprior}
\end{equation}
where $\Gamma_{A\rightarrow B}$ is a rotation matrix representing 
the rotation that occurs when a vector is parallel 
transported from tetrahedron A directly to tetrahedron B, $R_{tot}$ 
means the total rotation matrix after going through the 
whole path,  and $R1_B$ is $R1$ written in B's coordinate 
system; the other rotations, with no subscripts, are written in A's 
coordinate system.  Eq.~(\ref{eq:3dnullpath}) can be written completely 
in A's coordinate system as
\begin{equation}
1=R_{tot}=R1\; R4 \; R2 \; R3
\label{eq:3dnullrot}
\end{equation}

Also in reference to the $\Gamma$'s, there is a subtlety to note.
The $\Gamma$'s are always "direct" $\Gamma$'s between adjacent 
d-simplices.  For example, 
$\Gamma_{A\rightarrow B}$ is not $\Gamma_{A\rightarrow C\rightarrow B}$ (the latter being starting 
at A, then go directly to C, and then go directly to B); 
indeed, since 
\begin{equation}
R1=\Gamma_{A\rightarrow B}\Gamma_{A\rightarrow C\rightarrow B}^{-1}
\label{eq:Gammasubtle}
\end{equation}
both cannot simultaneously be set equal to the unit matrix.

In general, when setting up a d-dimensional global coordinate system, one 
specifies the coordinate system in one d-simplex and $N_d-1$ $\Gamma$'s 
($N_d$ is the number of d-simplices in the lattice) such 
that all d-simplices are linked directly or indirectly; to specify any more 
$\Gamma$'s would be to specify a pre-determined (via edge lengths) loop 
rotation.  One can, for e.g., choose all $N_d-1$ $\Gamma$'s equal to the unit 
matrix, so that $\Gamma_{A\rightarrow B}$ in Eq.~(\ref{eq:3dnullrotprior}) 
would be one of three $\Gamma$'s set equal to the unit matrix.

If one only specifies $\Gamma$'s between adjacent d-simplices, 
one can think of this coordinate 
system in flat d-space as d-simplices which 1) are attached to each other only 
by the (d-1)-simplex ``faces'' 
corresponding to the $N_d-1$ $\Gamma$'s, 
and 2) can go through each other (for e.g., consider a hinge with 
greater than $2\pi$ radians around it) .  Of course, when viewed in the actual 
curved space, the d-simplices do not go though each other.
Some sample coordinate systems would be obtained by 
1) specifying $\Gamma$'s between 
one specific tetrahedron (or d-simplex) and every other one, such 
as $\Gamma_{A\rightarrow B}$, $\Gamma_{A\rightarrow C}$ and 
$\Gamma_{A\rightarrow D}$ in our particular 
set up, and 
2) specifying the $\Gamma$'s in a chainlike fashion, such as $\Gamma_{A\rightarrow B}$, 
$\Gamma_{B\rightarrow C}$, and $\Gamma_{C\rightarrow D}$ in our set up.

\vskip 160pt

\begin{center}
\leavevmode
\epsfysize=7cm
\epsfbox{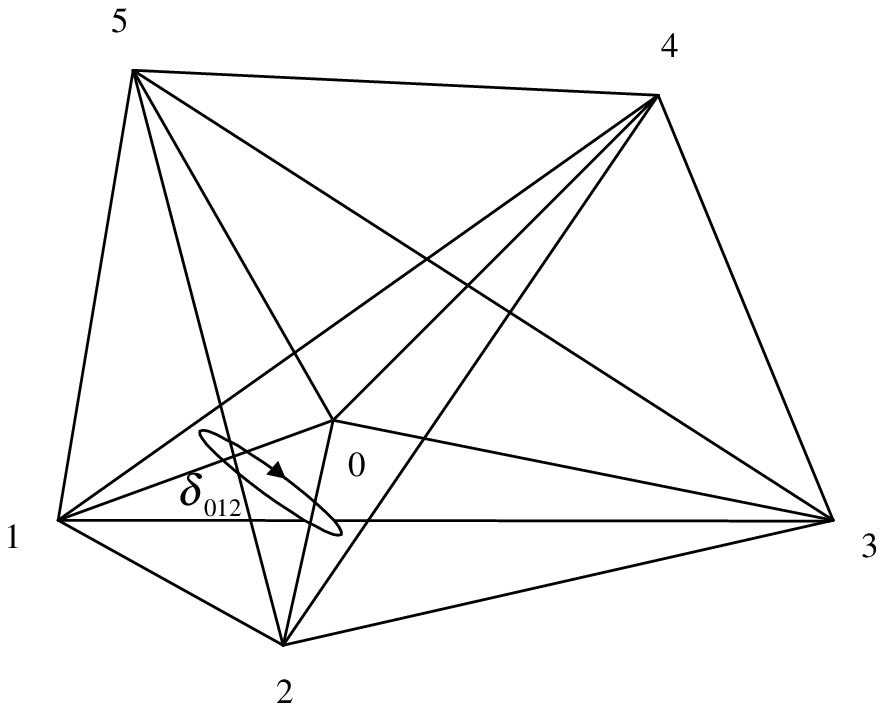}
\end{center}
\noindent
{\small{\it Fig.\ 4.
Similar to Fig.\ 1., but now in four dimensions.
In four dimensions five simplices meet on a point, labeled here by $0$.
Deficit angles $\delta_{012}$ etc. are associated with triangles $0-1-2$
etc. respectively.
The Bianchi identities are obtained by taking an ordered product of
rotation matrices along a path which encircles several hinges (triangles here)
and is topologically trivial, in the sense that it can be shrunk to a point.
\medskip}}

\vskip 20pt

Now, let us consider one four-dimensional set up in particular.
Consider a four 
simplex with a point in the middle.  This divides the original 
4-simplex into 5 new 4-simplices.  Then, readjust the edge lengths 
and curve the space (see Fig. 4).  We now consider parallel transporting 
the vector around a null path within this simplicial complex.  Label each 
of the new 4-simplices A, B, C, D and E (see Fig. 5).

\vskip 90pt

\begin{center}
\leavevmode
\epsfysize=7cm
\epsfbox{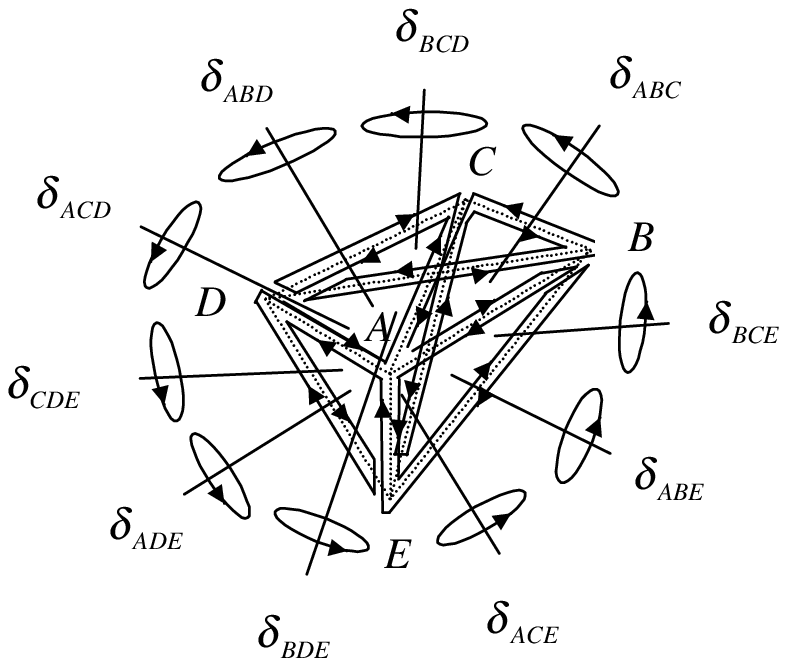}
\end{center}
\noindent
{\small{\it Fig.\ 5.
Similar to Fig.\ 2., but now in four dimensions.
The topologically trivial path which encircles all shown hinges
can be shrunk to a point.
The vertices $A$, $B$, $C$, $D$ and $E$ reside in the dual lattice,
and can conveniently be placed at the centers of the simplices
shown in Fig.\ 4..
As in the three-dimensional case, it is advantageous to label the
deficit angles by the vertices in the dual lattice.
The ordered product of rotation matrices around the shown path 
then reduces to the identity matrix.
\medskip}}

\vskip 20pt

We now make an important point for this particular set up:  
the number of 4-simplices around a triangle 
hinge is 3.  There are 3 vertices in the 
two dimensional space of the hinge and three vertices in the 
remaining two dimensions.  The three vertices in the hinge 
and any two of the three vertices outside the hinge 
form a 4-simplex, so that three 4-simplices surround the hinge.
\footnote{This three d-simplex result is now easily seen to be true 
in any number of dimensions for the point in the middle set up.}

\vskip 90pt

\begin{center}
\leavevmode
\epsfysize=7cm
\epsfbox{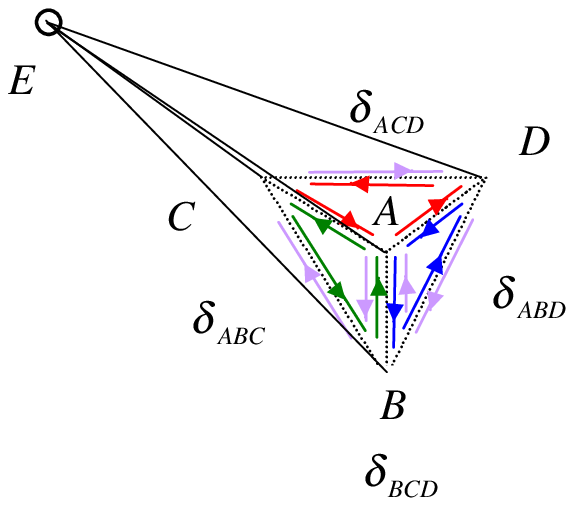}
\end{center}
\noindent
{\small{\it Fig.\ 6.
Similar to Fig.\ 3., but now in four dimensions.
Paths associated with different hinges
are shown here in different shades of gray.
The product of rotation matrices around a single hinge is simply 
related to the deficit angle for that hinge.
Again, in order to obtain a closed path segment AB has to be transversed
more than two times.
In constructing a product of rotations, it is sufficient to consider
only four of the five vertices; here, vertex $E$ is excluded.  In higher 
dimensional analogues, 
one has vertices F, G, etc., which are also not used.
\medskip}}

\vskip 20pt

So, we now consider a path which includes 4 of the 5 simplices 
which surround the "point in the middle".  Specifically, 
referring to Fig. 6, the path is
\begin{equation}
A\rightarrow B\rightarrow D\rightarrow A, A\rightarrow D\rightarrow C\rightarrow A, A\rightarrow C\rightarrow B\rightarrow A, A\rightarrow B, B\rightarrow C\rightarrow D\rightarrow B, B\rightarrow A
\label{eq:4dnullpath}
\end{equation}
The same path works for 3-d, which then includes all 4 of the 4 
tetrahedra surrounding the point in the middle (see Eq.~(\ref{eq:3dnullpath})).

\vskip 30pt
\newsection{Product of Rotations in Three Dimensions}
\hspace*{\parindent}

Having described the general geometric setup in the previous section,
we now proceed to give an explicit form for the rotation matrices in three
dimensions, and derive an exact form of the lattice Bianchi identity
by considering a product of rotation matrices along a homotypically
trivial path.
It turns out to be very convenient to be able to express 
a rotated vector in terms of the "old" vector and the hinge 
edge.  One notes that the only rotation occurs to the portion 
of the vector that is perpendicular to the hinge, so that 
$(\vec{v}\cdot\hat{l})\hat{l}$ is part of the new vector.  One proceeds 
to form an orthogonal coordinate system with the old vector,~$\vec{v}$, 
and the hinge, $\hat{l}$, using $\hat{l}$,$\vec{v}\times\hat{l}$ and 
$\hat{l}\times (\vec{v}\times\hat{l})$.  So, since $\vec{v}$ clearly has 
no component parallel to $\vec{v}\times\hat{l}$, the component of v 
which is rotated in the 
$\left[\vec{v}\times\hat{l}\right]$-$\left[\hat{l}\times (\vec{v}\times\hat{l})\right]$ 
plane is parallel to $\hat{l}\times (\vec{v}\times\hat{l})$.  So, one finds
\begin{equation}
\vec{v}^\prime = (\vec{v}\cdot\hat{l})\hat{l}+
\left\{\frac{\vec{v}\cdot [\hat{l}\times (\vec{v}\times\hat{l})]}
 {|\hat{l}\times (\vec{v}\times\hat{l})|}\right\}
\left\{\cos \delta \left[\frac{\hat{l}\times (\vec{v}\times\hat{l})}{|\hat{l}\times (\vec{v}\times\hat{l})|}\right] 
- \sin \delta \left[\frac{\vec{v}\times\hat{l}}{|\vec{v}\times\hat{l}|}\right]\right\} 
\label{eq:3dvprimeprelim}
\end{equation}
which simplifies to
\begin{equation}
\vec{v}^\prime = \left[ 2\sin^2\frac{\delta}{2} (\vec{v}\cdot\hat{l})\right]\hat{l}+(\cos\delta )\vec{v}+(\sin\delta )\hat{l}\times\vec{v}
\label{eq:3dprime}
\end{equation}
The total rotated vector (after all successive rotations) is most easily found using a recursive application 
of Eq.~(\ref{eq:3dprime}) via standard dot/cross product rules, as well as being sure to set up one's coordinate system 
where the $\Gamma_{A\rightarrow B}$ in Eq.~(\ref{eq:3dnullrotprior}) 
equals 1.  Then, one notes that the most general form of a rotation can be 
written as 
\begin{equation}
R_{tot}\vec{v}=a\vec{v}+\vec{b}\times\vec{v}+\sum_i(\vec{c}_i\cdot\vec{v})\vec{d}_i = 
\left( \begin{array}{lcr}
  a  + c_{i1}d_{i1} & -b_3 + c_{i2}d_{i1} &  b_2 + c_{i3}d_{i1} \\

 b_3 + c_{i1}d_{i2} &   a  + c_{i2}d_{i2} & -b_1 + c_{i3}d_{i2} \\

-b_2 + c_{i1}d_{i3} &  b_1 + c_{i2}d_{i3} &   a  + c_{i3}d_{i3}
\end{array} \right) \vec{v} 
\label{eq:3dprodeq}
\end{equation}
where $a$, $\vec{b}$, and the $\vec{c}_i$ are easily found with the expression one 
had gotten for $\vec{v}^{\prime}_{final}$ using Eq.~(\ref{eq:3dprime}).
Here, $\vec{c}_i$  runs over all possible vectors in the rotation, which for the 
sample case we are considering, is just $l_1,l_2,l_3,l_4$ and their 
$6$ cross products.  The $\vec{d}_i$ are then to be viewed 
as the vector coefficients of the $4+6=10$ $\vec{c}_i\cdot\vec{v}$'s in 
Eq.~(\ref{eq:3dprodeq}) (look at the non-matrix expression).  
The $a$, the $\vec{b}$ and the $\vec{d}_i$ are extremely 
complicated expressions in terms of deficit angles, internal 
angles, normalized volumes and normalized areas.  Here, normalization 
means dividing by the product of the magnitudes of the defining vectors.
Do note that due to the vector nature of the cross product in 3-d, 
$a$ does not contain normalized areas.

From the matrix form of Eq.~(\ref{eq:3dprodeq}) one can find a scalar and vector 
identity.  We find these equations by noting
\begin{equation}
R_{tot}-I \; = \; 0
\label{eq:Id3d}
\end{equation}
Taking the trace of both sides while using Eq.~(\ref{eq:3dprodeq}) gives the scalar equation
\begin{equation}
3(a-1) \, + \, \sum_i\vec{c}_i\cdot\vec{d}_i \; = \; 0
\label{eq:3dscalid}
\end{equation}
while taking the antisymmetric part of Eq.~(\ref{eq:3dprodeq}) gives the vector equation
\begin{equation}
-\vec{b}-\frac{1}{2}\sum_i\vec{c}_i\times\vec{d}_i \; = \; 0
\label{eq:3dvectid}
\end{equation}
and taking the symmetric part of Eq.~(\ref{eq:3dprodeq}) gives a tensor equation which 
is unnecessary to discuss.  
So, here, we have found a vector and scalar analogy of the product 
of rotations 
for arbitrary deficit angles.
These (un-contracted)
identities are in terms of deficit angles, internal triangle angles, 
normalized areas and normalized volumes.
The normalized areas, naturally vectors from a geometric point of view, appear explicitly only in 
the vector identity.  
Again, it should be noted that Eq.~(\ref{eq:Id3d}) is only valid for a product of rotations around 
hinges which gives no rotation.  Permute two of the rotation matrices making up the total rotation, 
and the new total rotation matrix will not be the unit matrix.

The explicit form of $R_{tot}$ in Eq.~(\ref{eq:3dprodeq}) can then
be used, in combination with Eq.~(\ref{eq:3dvectid}),
to write the 3-dimensional completely contracted Bianchi identities as
\begin{equation}
- \frac{1}{V(v)} \; \epsilon_{\alpha\beta\gamma}\epsilon^{\alpha\beta\rho}
(-\vec{b}-\frac{1}{2}\sum_i\vec{c}_i\times\vec{d}_i)_{\rho} \; = \; 0
\end{equation}
which simplifies to
\begin{equation}
\frac{1}{V(v)} \; (2\vec{b}+\sum_i\vec{c}_i\times\vec{d}_i)_{\gamma} \; = \; 0
\label{eq:3dfcbi}
\end{equation}
Let us now consider the small deficit angle limit to first order.  This 
simplifies Eq.~(\ref{eq:3dprime}) dramatically to 
\begin{equation}
\vec{v}^{\prime}_{{\rm small} \delta}=\vec{v}+\delta (\hat{l}\times\vec{v})
\label{eq:vpapprox}
\end{equation}
which gives, after applying successive rotations,
\begin{equation}
\begin{array}{l}
a=1 \\
\vec{b}=\sum_i\delta_i\hat{l}_i \\
c \; = \; 0
\end{array}
\label{eq:abcapprox}
\end{equation}
so that Eq.~(\ref{eq:3dvectid}) becomes
\begin{equation}
-2\vec{b}= 0
\label{eq:vecapprox}
\end{equation}
which, following a similar derivation to that of Eq.~(\ref{eq:3dfcbi}), allows 
the fully contracted Bianchi identities to be written as
\begin{equation}
\frac{2}{V(v)} \; \vec{b}_{\gamma} \; = \; 0
\label{eq:3dfcbismall}
\end{equation}
Eq.~(\ref{eq:3dscalid}) is a trivial 0=0 to first order.  To second order, 
it becomes
\begin{equation}
-\vec{b}\cdot\vec{b} \; = \; 0
\label{eq:scalapprox}
\end{equation}
where the $\vec{b}$ used is just the $\vec{b}$ in Eq.~(\ref{eq:abcapprox}).  
This equation can clearly be seen as a consequence of 
Eq.~(\ref{eq:vecapprox}), a first order equation.
Eqs.~(\ref{eq:vecapprox})~and~(\ref{eq:3dfcbismall}) (as well
as Eq.~(\ref{eq:ucbismall}), reduced to three dimenions)
have been verified to 1st order, and 
Eq.~(\ref{eq:scalapprox}) has been verified to 2nd order.
But for higher orders, 
these equations have been shown to be violated 
in favor of, repectively, 
Eqs.~(\ref{eq:3dvectid}), (\ref{eq:3dfcbi}), and
(\ref{eq:3dscalid}) (as well as Eq.~(\ref{eq:ucbirot}), again reduced
to three dimensions)
which are valid in all cases.
Additionally, all these equations are true for arbitrary set-ups,
not just our point in the middle set-up.

\vskip 30pt
\newsection{Product of Rotations in Four Dimensions}
\hspace*{\parindent}

The four-dimensional case is quite similar to the three-dimensional
case discussed in the previous section, and no major additional
complexities arise.
We will give here again an explicit form for the rotation matrices in four
dimensions, and derive an exact form of the four-dimensional
lattice Bianchi identity
by considering a product of rotation matrices along a homotypically
trivial path.
In the four dimensional case,
considering how a vector rotates when it is parallel 
transported around a hinge, we can, as in the three dimensional case,
form again an orthogonal 
coordinate system using the (old) vector and the hinge, 
the additional edge in the hinge compensating for going up one dimension.
Taking $\hat{l}_1$ and $\hat{l}_2$ as the two 
edges which form the hinge, one uses the following four vectors to form the orthogonal 
coordinate system:  $\hat{l}_1$, $\hat{l}_2^\prime$, $\vec{v}\times\hat{l}_1\times\hat{l}_2^\prime$ 
and $\hat{l}_2^\prime\times\hat{l}_1\times (\vec{v}\times\hat{l}_1\times\hat{l}_2^\prime)$ where

\begin{equation}
\hat{l}_2^\prime =\frac{\hat{l}_2-(\hat{l}_2\cdot\hat{l}_1)\hat{l}_1}
                      {|\hat{l}_2-(\hat{l}_2\cdot\hat{l}_1)\hat{l}_1|}
                 =\frac{\hat{l}_2-(\hat{l}_2\cdot\hat{l}_1)\hat{l}_1}
                       {2A[\hat{l}_1,\hat{l}_2]}
\label{eq:4dl2p}
\end{equation}
and where $A[\hat{l}_1,\hat{l}_2]$ is the area of the triangle, not the parallelogram, 
formed by $\hat{l}_1$ and $\hat{l}_2$.

Now, in four dimensions, the hinge we rotate about is a triangle, and the plane in which the 
rotation occurs is perpendicular to that triangle.  Since $\hat{l}_1$ and $\hat{l}_2^\prime$ span 
the space of the triangle, the components of the rotated vector in those two perpendicular 
directions will not change.  Now, since the only remaining component of the vector is in the 
$\hat{l}_2^\prime\times\hat{l}_1\times (\vec{v}\times\hat{l}_1\times\hat{l}_2^\prime)$ direction, 
v clearly being perpendicular to $\vec{v}\times\hat{l}_1\times\hat{l}_2$, we can write

\begin{equation}
\vec{v}^\prime = (\vec{v}\cdot\hat{l}_1)\hat{l}_1+(\vec{v}\cdot\hat{l}_2^\prime )\hat{l}_2^\prime 
               + \left\{ \frac{\vec{v}\cdot [\hat{l}_2^\prime\times\hat{l}_1\times (\vec{v}\times\hat{l}_1\times\hat{l}_2^\prime)]}
                                           {|\hat{l}_2^\prime\times\hat{l}_1\times (\vec{v}\times\hat{l}_1\times\hat{l}_2^\prime)|}
                 \right\}
                 \left\{ \cos\delta\left[ \frac{\hat{l}_2^\prime\times\hat{l}_1\times (\vec{v}\times\hat{l}_1\times\hat{l}_2^\prime)}
                                              {|\hat{l}_2^\prime\times\hat{l}_1\times (\vec{v}\times\hat{l}_1\times\hat{l}_2^\prime)|}
                                  \right]
                     +  \sin\delta\left[ \frac{\vec{v}\times\hat{l}_1\times\hat{l}_2^\prime}
                                            {|\vec{v}\times\hat{l}_1\times\hat{l}_2^\prime|}
                                  \right]
                 \right\}
\label{eq:4dvprimeprelim}
\end{equation}
which simplifies to

\begin{equation}
\vec{v}^{\; \prime} = 2\sin^2
\frac{\delta}{2}[(\vec{v}\cdot\hat{l}_1)\hat{l}_1+(\vec{v}\cdot\hat{l}_2^\prime )\hat{l}_2^\prime ]
+ [\cos\delta] \, \vec{v}
+[\sin\delta] \, \vec{v}\times\hat{l}_1\times\hat{l}_2^\prime
\label{eq:4dvprime}
\end{equation}
Now, as in the three dimensional case, we consider null paths as being products of rotations as we parallel transport 
a vector; only now we are going through 4-simplices as opposed to tetrahedra.  Recall that in the three dimensional case, 
the path could be written in terms of the four tetrahedra it went through, and only depended on the fact that each 
tetrahedron was directly connected to every other tetrahedron.  We can follow the same procedure in the four dimensional 
case by choosing four of the five 4-simplices and applying Eq.~(\ref{eq:4dnullpath}) (we could interchange 4-simplex E with any 
of the other 4-simplices in Eq.~(\ref{eq:4dnullpath})).  We have

\begin{equation}
R_{l_1,tot}-I=\prod_{m=1}^4R_{l_1,m}-I \; = \; 0
\label{eq:4dprodrot}
\end{equation}
where the $l_1$ is the same as in the previous equations.  
In general, the $l_1$ index can be taken to represent the ``hinge base'' for 
the product of rotations, i.e. the edge contained in all hinges involved in 
the product of rotations; the $l_2^\prime$ of the previous equations, on 
the other hand, varies from hinge to hinge.  
In our set-up, $l_1$ is the edge ``opposite'' the excluded 4-simplex.
(To be sure that one does not have the order 
of rotations for rotating by $+\delta$'s reversed, one can either project 
out along $l_1$ or simply try both orders).

Following the 3-d case, we write the most general product of matrices, and hence of rotation matrices, as

\begin{eqnarray}
R_{l_1,tot}\vec{v}  =  a\vec{v}+\sum_{n}\vec{\epsilon}^{\_}_{\mbox{ }\alpha\beta\gamma}b^\alpha_{1n}b^\beta_{2n}v^\gamma+\sum_i(\vec{c}_i\cdot\vec{v})\vec{d}_i =  a\vec{v}+\sum_{k}\sum_{j_k=1}^{(j_k)_{max}}\vec{\epsilon}^{\_}_{\mbox{ }\alpha\beta\gamma}b^\alpha_{1k}b^\beta_{2j_k}v^\gamma+\sum_i(\vec{c}_i\cdot\vec{v})\vec{d}_i \nonumber \\
              =    \left(\begin{array}{cccc}
    a+c_i^1d_i^1    & \sum_{kj}(b_{1k},b_{2j})^{34} + c_i^2d_i^1 & \sum_{kj}(b_{1k},b_{2j})^{42} + c_i^3d_i^1 & \sum_{kj}(b_{1k},b_{2j})^{23} + c_i^4d_i^1\\
\sum_{kj}(b_{1k},b_{2j})^{43} + c_i^1d_i^2 &     a+c_i^2d_i^2    & \sum_{kj}(b_{1k},b_{2j})^{14} + c_i^3d_i^2 & \sum_{kj}(b_{1k},b_{2j})^{31} + c_i^4d_i^2\\
\sum_{kj}(b_{1k},b_{2j})^{24} + c_i^1d_i^3 & \sum_{kj}(b_{1k},b_{2j})^{41} + c_i^2d_i^3 &     a+c_i^3d_i^3    & \sum_{kj}(b_{1k},b_{2j})^{12} + c_i^4d_i^3\\
\sum_{kj}(b_{1k},b_{2j})^{32} + c_i^1d_i^4 & \sum_{kj}(b_{1k},b_{2j})^{13} + c_i^2d_i^4 & \sum_{kj}(b_{1k},b_{2j})^{21} + c_i^3d_i^4 &     a+c_i^4d_i^4
\end{array}\right)\vec{v}
\label{eq:4dprodeq}
\end{eqnarray}
where
\begin{equation}
(b_{1k},b_{2j})^{\alpha\beta}\equiv b_{1k}^\alpha b_{2j}^\beta -
                                    b_{1k}^\beta b_{2j}^\alpha
\end{equation}
The main difference between Eq.~(\ref{eq:4dprodeq}) and Eq.~(\ref{eq:3dprodeq}) is that in Eq.~(\ref{eq:3dprodeq}) there was only one $\vec{b}$, whereas 
here, we have a set of pairs of $\vec{b}$'s, i.e. $\{(\vec{b}_{1n},\vec{b}_{2n})\}$.  So, for these new $\vec{b}$'s, the first lower 
index indicates whether it comes 1st or 2nd in the cross product with $\vec{v}$, the second lower index indicates what term in the sum 
it belongs to, and the upper index indicates which component of it is being taken.  
In our point in the middle set up, the $\vec{c}_i$'s consist of 5 edges 
and their 10 triple cross products; there are 15 corresponding $\vec{d}_i$'s.  
The $\sum_{n}$ form has a nicer appearance, but 
the $\sum_{kj}$ form is much better when doing computations.  The k subscript in $j_k$ has been omitted in the matrix form given.

Using Eq.~(\ref{eq:4dprodrot})
and taking the trace and the antisymmetric part of the equation give,
respectively,
\footnote{In $d$ dimensions, $4(1-a)\rightarrow d(1-a)$.}

\begin{equation}
4(a-1) \, + \, \sum_i\vec{c}_i\cdot\vec{d}_i \; = \; 0
\label{eq:4dscalidA}
\end{equation}
and

\begin{equation}
B \, - \, \frac{1}{2} \, C \; = \; 0
\label{eq:4dtensidA}
\end{equation}
where B and C have matrix elements\footnote{In $d$ dimensions, there are 
$d-2$ $b$'s contracted with the Levi-Civita tensor.}

\begin{equation}
B^{\alpha\beta} \equiv \sum_{jk}\epsilon^{\alpha\beta}_{\;\;\;\;\gamma\delta}b^\gamma _{1j}b^\delta _{2k} \nonumber \\
\label{eq:Beqn}
\end{equation}
and
\begin{equation}
C^{\alpha\beta} \equiv \sum_i(c^\alpha _i d^\beta _i - d^\alpha _i c^\beta_i) \nonumber \\
\label{eq:Ceqn}
\end{equation}
These (un-contracted)
identities are in terms of deficit angles, internal triangle angles, 
normalized areas and normalized four-volumes.  The normalized areas, naturally 
2-index tensors, appear explicitly only in the tensor identity, 
Eq.~(\ref{eq:4dtensidA}).

Using the explicit form of $R_{l_1,tot}$ in Eq.~(\ref{eq:4dprodeq}) 
and Eq.~(\ref{eq:4dtensidA}) the completely contracted Bianchi identities
can then be written as
\begin{equation}
-\frac{1}{2}\frac{1}{^4V(v)}\sum_{l_1\supset v}
\epsilon_{\alpha\beta\gamma\lambda} \, 
(2B-C)^{\alpha\beta}l_1^\lambda  \; = \; 0
\label{eq:4dfcbi}
\end{equation}
To first order in deficit angles, 
Eqs.~(\ref{eq:4dscalidA})~and~(\ref{eq:4dtensidA}) reduce to, 
respectively, $0=0$ and 
\begin{equation}
B^{\alpha\beta}=
\sum_i\delta_i \,
\epsilon^{\alpha\beta}_{\;\;\;\;\mu\nu} \, l_1^\mu l_{2i}^{\prime\nu}
\; = \; 0 
\label{eq:4dsmalleps}
\end{equation}
The above 
equation can be rewritten, 
using Eq.~(\ref{eq:bivector}), to give
\begin{equation}
\sum_i \delta_i \, U^{\alpha\beta}_i \; = \; 0 
\label{eq:U}
\end{equation}
where we have substituted in for $l_i^\prime$ using Eq.~(\ref{eq:4dl2p}).
Following a similar derivation to that of Eq.~(\ref{eq:4dfcbi}), the above 
equation allows the fully contracted Bianchi identities to be written 
for small angles as
\begin{equation}
-\frac{1}{^4V(v)}\sum_{l_1\supset v}\epsilon_{\alpha\beta\gamma\lambda}
\, (\sum_i \delta_i \, U_i^{\alpha\beta}) \, l_1^\lambda  \; = \; 0
\label{eq:4dfcbismalleps}
\end{equation}
Eqs.~(\ref{eq:U}),~(\ref{eq:4dfcbismalleps}) (as well as the
Eq.~(\ref{eq:ucbismall}), reduced to four dimensions)
have been verified to first order.
For higher order, 
these equations have been shown to be violated in favor of, respectively, 
Eqs.~(\ref{eq:4dtensidA}) and (\ref{eq:4dfcbi})
(as well as Eq.~(\ref{eq:ucbirot}), again reduced to four dimensions).
Also, Eq.~(\ref{eq:4dscalidA}), a trivial $0=0$ to first order, has been 
shown to be true generally.
\footnote{
We note here that these results apply to the Lorentz 
case as well.  One sets $ct=-ix_4$, and finds the $\vec{x},t$ rotations 
equivalent to the $\vec{x},x_4$ rotations.  Since all the Bianchi 
products of rotations, 
when written out in terms such as the $\Gamma_{A\rightarrow B}$ gauge 
transformation of Section 3, end up canceling to the unit matrix via 
products such as $\Gamma_{A\rightarrow B}\Gamma_{A\rightarrow B}^{-1}$, 
and as each such 
Euclidean gauge transformation has an equivalent Lorentz transformation, 
the Regge Calculus Bianchi identities hold in the Lorentz case as well.  
In finding these equivalent transformations, it is important to note that 
any rotation plane will have at least one spatial axis.  If the other 
(perpendicular) 
axis is also spatial, one has the standard rotation by the computed 
deficit angle.  If the other axis is time-like, one first computes the deficit 
angle as in the Euclidean case, multiplies it by $i$ to get the 
appropriate deficit angle to use in the sines and cosines, and ends up 
using hyperbolic sines and cosines.  For this case, it is important not 
to subtract out any multiple $2\pi$ from the original Euclidean deficit angle 
before it is multiplied by $i$.  If the other 
axis is null, there is no rotation, no matter what the deficit angle would 
be in the Euclidean case, as a null vector and a spatial vector cannot 
rotate into each other.  These three cases are, respectively, equivalent to 
$A^{\mu\nu}_\Gamma A_{\mu\nu}^\Gamma >0$, 
$A^{\mu\nu}_\Gamma A_{\mu\nu}^\Gamma <0$ and 
$A^{\mu\nu}_\Gamma A_{\mu\nu}^\Gamma =0$; alternatively, they are also 
equivalent to, respectively, 
$A^{\gamma\delta_1...\delta_{d-3}}_hA_{\gamma\delta_1...\delta_{d-3}}^h <0$, 
$A^{\gamma\delta_1...\delta_{d-3}}_hA_{\gamma\delta_1...\delta_{d-3}}^h >0$ 
and 
$A^{\gamma\delta_1...\delta_{d-3}}_hA_{\gamma\delta_1...\delta_{d-3}}^h =0$.
Lastly, it is important to note that, once one fixes a going around coordinate 
system, one needs Euler angles to write a space-time rotation.}

\vskip 30pt
\newsection{Discrete Riemann Tensor and its Dual}
\hspace*{\parindent}

In the previous two sections an explicit, exact form of the lattice Bianchi
identity was derived by considering a product of rotation matrices
(in three and four dimensions respectively) along a homotypically
trivial path.
Curvature enters the above lattice Bianchi identities through the deficit angle,
but one interesting question left partially open is the relationship
between the exact lattice Bianchi identities (which involve products of
rotation matrices) and the continuum Bianchi identities (which
involve derivatives of the Riemann tensor).
In this and the following section an expression of the Riemann
tensor and its dual will be derived, which will eventually be used
to show that in four dimensions (Section 9), if one proceeds from
the continuum Bianchi identities (in integrated form) and
inserts the expression for the discrete Riemann tensor discussed
below one obtains, after promoting infinitesimal rotations to finite rotations,
the same exact lattice Bianchi identity discussed previously
in Section 4 (in Section 10 the same set of result is presented
in three dimension).

Having obtained the necessary ingredients to describe arbitrary
lattice rotations of
vectors, we proceed next to derive a general form for the lattice
Riemann curvature tensor.
An explicit form for the Riemmann tensor in term of rotations will be
quite useful here, since it will allow us to establish a relationship
(known to exist in the continuum) between the Regge field equations
and the lattice Bianchi identities derived in the previous sections
in three and four dimensions.
Furthermore it will allow us to check the overall consistency of our results,
since later in the paper we will re-derive the lattice Bianchi identities
by starting from their continuum expression in terms of the
Riemann tensor, and by promoting infinitesimal rotations to
finite rotations will be able to show that the resulting lattice Bianchi
identities are in fact identical (in d=3 and d=4) to the expressions
previously derived in the preceding sections. 

Consider moving a vector $V$ once around a Voronoi loop, i.e. a loop formed by Voronoi edges 
surrounding a hinge.  The change in $V$, denoted here by $\delta V$, is then 
given by

\begin{equation}
\delta V^\alpha = ({\mathbf{R-1}})^{\alpha}_{\;\;\beta}V^\beta
\end{equation}
where ${\mathbf{R}}$ is the is the rotation matrix associated with the hinge.  
Now, $\delta V$ in the continuum is given by

\begin{equation}
\delta V^\alpha = \half \; R^{\alpha}_{\;\;\beta\mu\nu} \; A^{\mu\nu}_{\Gamma} \; V^\beta
\label{eq:deltaV}
\end{equation}
where $A^{\mu\nu}_{\Gamma}$ is the antisymmetric bivector representing the loop area.
So we make the tentative identification

\begin{equation}
\half \; R^{\alpha}_{\;\;\beta\mu\nu}A^{\mu\nu}_{\Gamma} \; = \;
({\mathbf{R-1}})^{\alpha}_{\;\;\beta}
\label{eq:riemarearot}
\end{equation}

Now, because of the sums on $\mu$ and $\nu$, it is not immediately clear how 
one can divide by $A^{\mu\nu}_\Gamma$ to solve for the Riemann curvature 
tensor.  Let us, then, simply take a frame where there are only two components 
of $A^{\mu\nu}_{\Gamma}$, namely $A^{12}_{\Gamma}$ and 
$A^{21}_{\Gamma}$.  In that frame, using the antisymmetry of Riemann in 
$\mu$ and $\nu$, one can divide, and one finds

\begin{equation}
R^{\alpha}_{\;\;\beta 12}=\frac{({\mathbf{R-1}})^{\alpha}_{\;\;\beta}}
                               {A^{12}_{\Gamma}}
\end{equation}
Multiplying by unity in the form of $A_{12}^{\Gamma}$ (the $_{12}$ component 
of the index-lowered version of $A^{12}_{\Gamma}$) over itself, one finds

\begin{equation}
R^{\alpha}_{\;\;\beta 12}=
\frac{({\mathbf{R-1}})^{\alpha}_{\;\;\beta}A_{12}^{\Gamma}}
     {A_{\Gamma}^2}
\label{eq:riemspecial}
\end{equation}
where the area of the loop is an invariant satisfying 
$A_{\Gamma}^2=A_{\lambda\kappa}^{\Gamma}A^{\lambda\kappa}_{\Gamma}/2$. 
Of course, all values of the Riemann curvature tensor with $\alpha$ or 
$\beta$ neither $1$ or $2$ are zero in this coordinate 
system.

Now, the above equation can be rewritten in this coordinate system as

\begin{equation}
R^{\alpha}_{\;\;\beta\mu\nu}=
\frac{({\mathbf{R-1}})^{\alpha}_{\;\;\beta}A_{\mu\nu}^{\Gamma}}
     {A_{\Gamma}^2}
\label{eq:riemfirst}
\end{equation}
However, this is now a tensor equation, and hence valid in all 
coordinate systems.  Still, there is a basic problem with this, 
and that is that $({\mathbf{R-1}})_{\sigma\beta}$ is only antisymmetric 
to first order (for example, $({\mathbf{R-1}})_{11}=0$ only to first order), 
so that this identification of the right hand side of 
Eq.~(\ref{eq:riemfirst}) with 
the Riemann curvature tensor and all of its symmetries can only be valid 
to first order.  So, let us write Eq.~(\ref{eq:riemfirst}) to first order 
in the deficit angle $\delta$:

\begin{equation}
R^{\alpha}_{\;\;\beta\mu\nu}=
\frac{\delta}{A_\Gamma }
U^{\alpha}_{\;\;\beta}
\frac{A_{\mu\nu}^{\Gamma}}
     {A_{\Gamma }}
\label{eq:riem}
\end{equation}
where, in our coordinate system, which we now take to be 
orthonormal,\footnote{A minus sign appears above in the Lorentz case of 
a x-t loop.} 
\begin{equation}
U_{\alpha\beta}=
\left(
\begin{array}{crcc}
 0 & -1 & 0 & 0 \\
 1 &  0 & 0 & 0 \\
 0 &  0 & 0 & 0 \\
 0 &  0 & 0 & 0
\end{array}
\right)
\label{eq:defmatrix12}
\end{equation}
The general form of $U_{\alpha\beta}$,  in an arbitrary coordinate system, is

\begin{equation}
U_{\alpha\beta}=
   \epsilon_{\alpha\beta\mu\nu}l_1^{\mu}l_2^{\nu}/2A_h 
\label{eq:defmatrix}
\end{equation}
where $l_1$ and $l_2$ are the two hinge vectors and 
$A_h=\sqrt{A_{\lambda\kappa}^h A^{\lambda\kappa}_h /2}$
with $A_h$ the area of the 
hinge and $A^{\lambda\kappa}_h$ the associated 
bivector.\footnote{In the Lorentzian case of a space-time hinge, 
$A_h=\sqrt{-A_{\lambda\kappa}^h A^{\lambda\kappa}_h /2}$.}
Note that $\delta$ is invariant because 
it is formed 
from angles which are arc cosines of ratios of invariant dot products 
to invariant lengths, and furthermore that 
$\epsilon_{\alpha\beta\mu\nu}=\sqrt{g} \; [1234]$ with $[1234]=+1$. 
So Eq.~(\ref{eq:defmatrix}) is a tensor, and hence so is Eq.~(\ref{eq:riem}). 
Since the expression in Eq.~(\ref{eq:riem}), after lowering the first index, also 
has the appropriate antisymmetry in the first two indices, 
we can see it as a Riemann tensor candidate.

Now, by going back to our special coordinate system, it can be easily 
verified that the dual of the normalized loop area is the normalized hinge area, i.e.

\begin{equation}
\frac{A_{\alpha\beta}^{\Gamma}}{A_\Gamma }=
                               \frac{\epsilon_{\alpha\beta\gamma\delta}}{2}
                               \frac{A^{\gamma\delta}_h}{A_h} 
\label{eq:dual}
\end{equation}
This allows us to rewrite Eq.~(\ref{eq:riem}) simply as
\begin{equation}
R^{\alpha}_{\;\;\beta\mu\nu}=
\frac{\delta}{A_\Gamma } \; U^{\alpha}_{\;\;\beta} U_{\mu\nu}
\label{eq:riemU}
\end{equation}

This expression has been discussed by the authors of Ref.\cite{hw84,lesh}, and is known to 
possess all requisite symmetry properties of the Riemann tensor.
However, as these authors note, it implies that the square of the Ricci scalar, 
the square of the Ricci tensor and the square of the Riemann tensor are 
all directly proportional, with constants of proportionality independent 
of the edge lengths.  
We now define
\begin{equation}
R^{\alpha}_{\;\;\beta\mu\nu}\equiv
R^{\alpha}_{\;\;\beta\mu\nu} (b,v) \equiv
\sum_{^{h\subset b}_{h\supset v}} R^{\alpha}_{\;\;\beta\mu\nu} (h)
\label{eq:Riemdef}
\end{equation}
for a formulation of Regge Calculus which breaks up space-time into 
d-boxes.  Here, one chooses one particular vertex $v$ to be 
the origin for the $g_{\mu\nu}$ of d-box $b$.
Then the metric can, for example, 
be defined via dot products of the box's non-diagonal edges containing $v$, 
or, equivalently, can be defined via edges in the box containing $v$ which 
are in one of the box's d-simplices.  Other vertices $v$ for other d-boxes 
are chosen to be in the same relative location within the d-boxes, so that 
each hinge curvature term $R^{\alpha}_{\;\;\beta\mu\nu} (h) $ is included once.

Two other definitions of the Riemann tensor are possible in principle, but
as we will be show below, fail to pass a crucial test.
Noting that each rotation plane has four hinges for a d-box, 
one could have alternately defined
\begin{equation}
R^{\alpha}_{\;\;\beta\mu\nu}\equiv
\frac{1}{4}\sum_{h\subset s} R^{\alpha}_{\;\;\beta\mu\nu} (h)
\label{eq:Riemdefboxavg}
\end{equation}
This could be useful in a situation with large curvatures 
where the increased computation due to averaging and, as we will see, 
more complex Einstein tensors $G_{\gamma}^{\;\;\gamma^\prime}$'s, is more than canceled out 
by being able to use fewer d-boxes due to the relaxing of the 
origin vertex location constraint.

Alternatively, if one uses only d-simplices and does not require them to form d-boxes, 
then one can define
\begin{equation}
R^{\alpha}_{\;\;\beta\mu\nu} \equiv
R^{\alpha}_{\;\;\beta\mu\nu} (s,v) \equiv
\sum_{h \subset s} \frac{1}{N_{s_h}} R^{\alpha}_{\;\;\beta\mu\nu} (h)
\label{eq:Riemdefsim}
\end{equation}
where $N_{s_h}$ is the number of d-simplices containing hinge 
$h$.

Now, one objection to these definitions is that the number of independent 
curvature components used per d-volume, for these three cases respectively
(Eqs.~(\ref{eq:Riemdef}), (\ref{eq:Riemdefboxavg}) and (\ref{eq:Riemdefsim})),
is $d(d-1)/2$, $2d(d-1)$ and $d(d+1)/2$, none of which is the expected $d^2(d^2-1)/12$, the 
number of components of Riemann in the continuum.  However, let us now 
calculate the number of components of Riemann about a vertex in the 
d-box case where consistent labeling of the origins are possible, i.e. 
where $2^d$ d-boxes surround each vertex.  
The number of m-boxes 
surrounding a given vertex is $2^m(\stackrel{d}{m})$ where we choose 
$m$ of the $d$ axis directions, and note that there are two choices per 
direction.  Setting $m=d-2$ gives 
\begin{eqnarray}
N_{h,v}=2^{d-3}d(d-1) & =d^2(d^2-1)/12 & d=2,3 \\
                      & >d^2(d^2-1)/12 & d\ge 4
\end{eqnarray}
So, if one
sets up a going around coordinate system at 
a vertex, and defines the curvature via
\begin{equation}
R^{\alpha}_{\;\;\beta\gamma\delta}\equiv\sum_{h\supset v}
R^{\alpha}_{\;\;\beta\gamma\delta} (h)
\label{eq:Riemregge}
\end{equation}
the symmetries of the form of the 
hinge curvatures keep the number of independent components of the 
curvature equal to $d^2(d^2-1)/12$ in any dimension.  This equation, 
then, having along with the natural form for the Riemann curvature tensor, 
all the symmetry properties of the continuum 
Riemann curvature tensor as well as the correct number of independent 
components, is the Regge analog of the Riemann curvature 
tensor.
\footnote{
Were we to use d-simplices which need not form d-boxes, then the 
consistent choice of origin vertices may be either impossible or very 
difficult.  
One might also make this choice for a d-box situation with high curvatures 
where the increased computation due to averaging is more than canceled out 
by being able to use fewer d-boxes.
But this is more complicated and our results for 
$G_{\gamma}^{\;\;\gamma^\prime}$ will also show a distinct preference for the 
first option.  Still, in cases where curvature is sufficiently high and 
computing power will not allow for a sufficient breakdown of space-time 
into d-boxes such that a consistent choice of origin vertices is possible, 
this definition may be useful.  Indeed, for this reason, were one to break down 
space-time into d-simplices without forming d-boxes, this definition would 
likely be necessary.
Still, in such a case where either a quick estimate is needed or 
where computing power is insufficient to use the first definition, this 
definition would then be useful. }

We are now in a position to calculate the {\it dual} Riemann tensor via

\begin{equation}
\tilde{R}^{\alpha\beta\gamma\delta}\equiv
\half \epsilon^{\mu\nu\gamma\delta}R^{\alpha\beta}_{\;\;\;\;\mu\nu} =
\half \delta \, \epsilon^{\mu\nu\gamma\delta}
U^{\alpha\beta}A_{\mu\nu}^{\Gamma}/
A_{\Gamma}^2
\label{eq:RtildR}
\end{equation}
where the loop now used is one of the aforementioned inner-d-box 
loops.\footnote{In the Lorentzian case, there is a minus sign in this 
definition of $\tilde{R}^{\alpha\beta\gamma\delta}$.} 
We can interpret the above result in the following way.
Here $U^{\alpha}_{\;\;\beta}$ 
corresponds to the ``effective'' $U^{\alpha}_{\;\;\beta}$ of the loop, 
and $\delta$ is the ``effective'' deficit angle of the loop.  
The ``effective'' Riemann curvature tensor is the sum of all 
the ``old'' Riemann curvature tensors corresponding to the areas 
that the $2^d-1=15$ edges form, so that the effective $\delta$ and 
$U^{\alpha\beta}$ 
are determined for each standard square loop (which can only encircle 
areas formed by the 15 edges).  Note that, since our ``old'' Riemann 
curvature tensors had all the symmetry properties of the continuum 
Riemann curvature tensor, and because our ``new'' (i.e. actual) Riemann 
curvature tensor is the sum of these old tensors, it too has all the 
symmetry properties of the continuum Riemann curvature tensor.  And, further, 
it has the advantage of allowing us to consider, in the case of small 
deficit angles, only perpendicular loops 
(more precisely, perpendicular loop areas), as we do in the continuum.
Now, using Eq.~(\ref{eq:dual}), 
Eq.~(\ref{eq:RtildR}) can be written as

\begin{equation}
\tilde{R}^{\alpha\beta\gamma\delta}=\frac{\delta}{A_\Gamma }
                                    U^{\alpha\beta}
                                    \frac{A^{\gamma\delta}_h}{A_h}
\label{eq:RtildA}
\end{equation}
which, incidentally, holds in any number of dimensions simply by adding 
$d-4$ extra indices after the $\delta$ indices.  
As the action will involve an integration, it will now be helpful 
to find out the total four-volume enclosed by the hinge and its loop 
in terms of the product of the loop area and the hinge area.

In the 4-d case the denominator, $A_\Gamma A_h$, is six times the four-volume 
enclosed by the hinge and the path, here denoted by $^4V(h,\Gamma )$, this 
latter four-volume being formed via the 
Voronoi vertices surrounding the hinge and the hinge vertices.  To see this 
fact, note that, in the 3-d case, as one approaches a 
vertex, the area perpendicular to the hinge is similar to the loop area, 
but gets smaller proportionally to the square of the perpendicular 
distance from the point,
labeled $p$, which marks the intersection of the loop area and the hinge.  
(In particular, all lengths defining the area are proportional to this 
distance, and hence areas, as functions of products of pairs of lengths, 
must be proportional 
to the square of this distance.)  
By letting $l_{1a}$ be the 
distance from one of $l_1$'s vertices to $p$, and letting $l_{1b}$ be 
the distance from $l_1$'s other vertex to $p$, one finds the total 
3-d volume formed by the loop and $l_1$ to be 
\begin{equation}
^3V=\int_0^{l_{1a}} (A_\Gamma )(s/{l_{1a}})^2ds+
    \int_0^{l_{1b}} (A_\Gamma )(s/{l_{1b}})^2ds
   =A_\Gamma l_1/3 
\label{eq:3V}
\end{equation}
In 4-d, the total four-volume formed by the loop and the hinge is obtained by 
noting that the three volume shrinks proportionally to the cube of the 
perpendicular distance from the three volume to the remaining vertex, so that, letting 
$l_2^\delta$ be the component of the remaining hinge edge 
which is perpendicular to the tetrahedron,

\begin{equation}
^4V=\int_0^{l_2^\delta }(A_\Gamma l_1/3)(s/l_2^\delta )^3ds
   =\epsilon_{\mu\nu\gamma\delta}A^{\mu\nu}_{\Gamma}l_1^\gamma l_2^\delta /24
   =\epsilon_{\mu\nu\gamma\delta}A^{\mu\nu}_{\Gamma}A^{\gamma\delta}_h /24
   =A_\Gamma A_h/6
\end{equation}
where the next to last term is found by taking our special coordinate system, 
with two axes along $A_\Gamma$ and two axes along $A_h$.
In d-dimensions, one has $^dV=\epsilon_{\mu\nu\gamma\delta ...}
                              A^{\mu\nu}_{\Gamma}V^{\gamma\delta ...}_h/d!
                              =2A_\Gamma V_h/d(d-1)$.
Incidentally, the only constraint on the loop here is that it be perpendicular 
to the hinge. 
One has therefore

\begin{equation}
\tilde{R}^{\alpha\beta\gamma\delta}=\frac{1}{6} \; 
                                    \frac{\delta}{^4V_h} \;
                                    U^{\alpha\beta}
                                    A^{\gamma\delta}_h
\label{eq:dualRReg}
\end{equation}
Using Eq.~(\ref{eq:riemU}), we find 
\begin{equation}
R=R_{\alpha\beta}^{\;\;\;\;\alpha\beta}=
\frac{\delta}{A_\Gamma}
\frac{\epsilon_{\alpha\beta\gamma_1...\gamma_{d-2}}
      A_h^{\gamma_1...\gamma_{d-2}}
      \epsilon^{\alpha\beta\gamma_1^{\prime}...\gamma_{d-2}^{\prime}}
      A^h_{\gamma_1^{\prime}...\gamma_{d-2}^{\prime}}}
     {[(d-2)!A_h]^2}
= 2 \frac{\delta}{A_\Gamma}
\label{eq:Riemscal}
\end{equation}
so that the Einstein-Hilbert action becomes
\begin{equation}
I=-\int_XR\sqrt{g} \; d^dx=-\sum_h\int_h\frac{2\delta}{A_\Gamma}\sqrt{g} \; d^dx
\label{eq:action}
\end{equation}
where $X$ is all space and $\int_h$ is defined as the integral over 
$^dV(h,\Gamma_h)$, the d-volume formed by the hinge and the loop area.
As the integrand is constant over the hinge, we do the integral by simply 
calculating this aforementioned d-volume.  
One can now can write the action as
\begin{equation}
I=-\frac{4}{d(d-1)}\sum_h \delta_h \, V_h
\label{eq:acthinge}
\end{equation}
where $ V_h / d(d-1)$ is the (d-2)-volume of the hinge, reducing to an area $A_h$ in 4-d.
(The sign in the above equation becomes positive in the Lorentz case.)

\vskip 30pt
\newsection{Action Variation and Einstein Tensor}
\hspace*{\parindent}

This section will be devoted to discussing the relationship between
the expression for the dual of the Riemann tensor, as given in the 
previos section, and the Regge field equations.
We will show that the above construction is indeed consistent, by
deriving from it the Regge field equations.
While this does not constitute a general proof of correctness of
the proposed expression, it does provide one rather significant
test.

Next one would like to vary the action with respect to $g_{\mu\nu}$.  
In the continuum each $g_{\mu\nu}$ is an independent variable 
upon which the geometry is based, which is why one varies the action with 
respect to them.  
Since, in Regge Calculus, the 
geometry is completely based on edge lengths, only edge lengths can be 
used in computing $g_{\mu\nu}$.  In particular, any $g_{\mu\nu}$ for 
the loop plane is not a true $g_{\mu\nu}$, since these quantities can be 
defined independent of edge lengths.  Furthermore, one can see that if one did 
vary the above action with respect to such a quantity, one would get zero 
because the hinge areas are independent of any quantity which is independent 
of the edge lengths.  
So, one is led to consider the metric $g_{\mu\nu}$ as being 
defined as a function of hinge edge lengths.  Here, we take 
one $(d-2)\times (d-2)$ $g_{\mu\nu}$ matrix per hinge.  
Using the invariance of $\delta$ in any 
dimension\cite{regge}, $^dV_h$ as the Voronoi d-volume surrounding the hinge 
and $V_h=\int_h\sqrt{|g|}d^{d-2}x/(d-2)!$, we now compute, 
for an individual hinge, 
\begin{eqnarray}
{\mathcal{G}}_{h\;\gamma}^{\;\;\;\;\;\gamma^\prime}
& \equiv &
\int_{^dV_h}G_{h\;\gamma}^{\;\;\;\;\;\gamma^\prime}\sqrt{|g|} \; d^dx
\; = \; -\frac{4}{d(d-1)} \, g_{\gamma\rho} \,
\delta_h \, \frac{\partial V_h}{\partial g_{\rho\gamma^\prime}} \nonumber \\
& = &
-\int_h\frac{4}{d!} \, \delta_h \,
\frac{g_{\gamma\rho}}{2\sqrt{|g|}} \, [\sigma_1...\sigma_{d-2}] \,
g_{\sigma_11}...g_{\sigma_{\gamma^\prime -1}\gamma^\prime -1}\delta^{\;\;\rho}_{\sigma_{\gamma^\prime}}
g_{\sigma_{\gamma^\prime +1}\gamma^\prime +1}...g_{\sigma_{d-2}d-2}
\, d^{d-2}x
\nonumber \\
& = & -\frac{2}{d(d-1)}\delta^{\;\;\gamma^\prime}_{\gamma}\delta_h V_h
\end{eqnarray}
To validate this result,\footnote{The sign is opposite for a space-time 
loop, because Eq.~(\ref{eq:acthinge}) changes sign for the reason 
given in the footnote for Eq.~(\ref{eq:riem}).  Also, letting 
$\sqrt{g}\rightarrow\sqrt{-g}$ for space-time hinges has no effect on 
the sign.}
which confirms the correctness of using $(d-2)\times (d-2)$ 
hinge metrics, we compute the above quantity in a different manner 
via\footnote{Once again, the sign of the result changes for a space-time 
loop for the reason given in the footnote to Eq.~(\ref{eq:riem}).  
For any Lorentzian hinge, 
the minus sign referenced in the footnote to Eq.(\ref{eq:RtildR}) is canceled 
by the minus sign in the inverse to Eq.~(\ref{eq:dual}).}
\begin{eqnarray}
{\mathcal{G}}_{h\;\gamma}^{\;\;\;\;\;\gamma^\prime}
&=&\int_{^dV_h}
G^{\;\;\;\;\;\gamma^\prime}_{h\;\gamma}\sqrt{|g|} \; d^dx
=-\frac{1}{2(d-3)!}\int_{^dV_h}
\epsilon_{\alpha\beta\gamma^\prime\delta_1...\delta_{d-3}}
\tilde{R}_h^{\alpha\beta\gamma\delta_1...\delta_{d-3}}\sqrt{|g|} \; d^dx 
\nonumber \\
&=&-\frac{1}{2(d-3)!}\frac{^dV_h}{A_\Gamma V_h}
\epsilon_{\alpha\beta\gamma^\prime\delta_1...\delta_{d-3}}
\delta_h U^{\alpha\beta}V_h^{\gamma\delta_1...\delta_{d-3}}
=-\frac{2}{d(d-1)}\delta_{\gamma}^{\;\;\gamma^\prime}\delta_h V_h
\end{eqnarray}
Now, noting that this final result is independent of the metric, 
and noting that the $\delta^{\;\;\gamma^\prime}_{\gamma}$ indicates that the 
components of ${\mathcal{G}}^{\;\;\gamma^\prime}_{\gamma}$ 
are most naturally taken along edges, 
we find
\begin{equation}
{\mathcal{G}}^{\;\; l^\prime}_{l}=-\frac{2}{d(d-1)}
\sum_{^{h\supset l}_{h\subset{\rm d-box}}}\delta_h V_h
\delta^{\;\; l^\prime}_{l}
\label{eq:Gll}
\end{equation}
In a given d-box of our global coordinate system, where these $l$'s are the 
box's (non-diagonal) axis $l$, the integrated Einstein tensor is defined 
via hinges within that d-box, so that there is no global over-counting 
of each contribution to ${\mathcal{G}}^{\;\; l}_{l}$.  We now easily see 
that, using our d-box's $g_{\mu\nu}$ defined by the axes' edges, 
\begin{eqnarray}
{\mathcal{G}}_{ll^\prime}=-\frac{2}{d(d-1)}
\sum_{^{h\supset l}_{h\subset{\rm d-box}}}\delta_h V_h
g_{ll^\prime}
& \mbox{and} &
{\mathcal{G}}^{ll^\prime}=-\frac{2}{d(d-1)}
\sum_{^{h\supset l}_{h\subset{\rm d-box}}}\delta_h V_h
\delta^{ll^\prime}
\label{eq:Gdduu}
\end{eqnarray}
The cosmological constant term is derived in each d-box from the 
d-box's $g_{\mu\nu}$ via the variation of $2\int\lambda\sqrt{g} \; d^dx$ 
exactly as it is in the continuum to get 
\begin{eqnarray}
\lambda g^{ll^\prime} & \mbox{and hence} & \lambda \delta^{\;\;\l^\prime}_{l}
\;\; \mbox{and} \;\;  \lambda g_{ll^\prime}  
\label{eq:cosconst}
\end{eqnarray}
This is the most natural result for the left hand side of Einstein's 
equations for Regge Calculus because of its metric independence.  
A computation of ${\mathcal{G}}^{\mu\nu}$ or ${\mathcal{G}}_{\mu\nu}$ for edges would have been 
counter-intuitive because one would be using a different metric for each 
hinge in the sum.  Also, using 
$G_{h\;\gamma}^{\;\;\;\;\;\gamma^\prime}$ 
would have explicitly involved the areas of the hinges' loops, 
which are not uniquely 
defined.
Now, the standard result for the field equations in four dimensions
\begin{equation}
\frac{l}{2}\sum_{h\supset l} \delta_h\cot\theta_{op}
\label{eq:Regeom}
\end{equation}
is derived by varying the integrated action with respect to the edge lengths, 
and hence is 
most similar to $\mathcal{G}^{\mu\nu}$, 
which is found by varying with respect to 
$g_{\mu\nu}$, which is a function of squares of edge lengths.  Hence 
this standard result is not metric independent.  However, as long as one 
has $T^{\mu\nu}=0$, there is no 
problem with it as Regge Calculus then only involves edge lengths.  However, 
if one wishes to include $T^{\mu\nu}$, \textit{and} if this 
$T^{\mu\nu}$ is not derivable from an action, then one must 
use Eq.~(\ref{eq:Gll}).  Also, Eq.~(\ref{eq:Gll}) seems more natural, 
and, except for having a more limited sum, has 
the same form as the action.
Lastly, we note that the individual terms in 
Eqs.(\ref{eq:Gll})~and~(\ref{eq:Regeom}) can be related via 
the transformations $g_{11}=l_1^2$, $g_{22}=l_2^2$ and 
$g_{12}=(l_1^2+l_2^2-L^2)/2$ so that the variation of the integral 
of the action can be transformed, for any one fixed term, 
between the two sets of variables via a Jacobian.

Ideally, one wants a one-to-one mapping between the (independent) edge length 
variables and the independent $g_{\mu\nu}$ variables.  This is easily achieved 
by choosing as the independent $g_{\mu\nu}$'s the $g_{\mu\mu}$'s, i.e. 
the metric components equal to the square of a given edge length.
Of course, we can always switch global metric variables by replacing 
some $g_{\mu\mu}$'s with an equal number of $g_{\mu\nu}$'s (while being sure 
to maintain the relative independence of the new set of metric components).  
We do this now, choosing $g_{11}=l_1^2$ for all $(d-2)\times (d-2)$ metrics 
for the hinges bordering $l_1$; we let all such metrics be based at one of 
$l_1$'s vertices.  All other lattice $g_{\mu\nu}$'s are then chosen such that 
the set of all $g_{\mu\nu}$'s are independent.  A simple way to make such 
a choice is to choose the metric components equal to the edge length squared 
of each edge not in a simplex bordering $l_1$.
Now, replacing $l_1$ with $L$ and varying with respect to $g_{11}=L^2$ gives, 
for 4-d
\begin{equation}
-\frac{1}{12}\sum_{h\supset L}\delta_h\cot\theta_{op}
\end{equation}
which is easily related to the standard result by multiplying by $-6L$.

As an exercise we show next the equivalence of our Einstein tensor to the 
variation of the action with respect to the edge lengths 
in the case of four dimensions.
For a hinge with edges $l_1$ and $l_2$, taking 
$g_{11}=l_1^2$, $g_{22}=l_2^2$ and $g_{12}=l_1\cdot l_2=(l_1^2+l_2^2-L^2)/2$ 
where $L$ is the third edge of the triangle formed by $l_1$ and $l_2$, 
we have
\begin{equation}
A=\frac{1}{2}\sqrt{g_{11}g_{22}-g_{12}g_{21}}
\end{equation}
where, for variation purposes, we take $g_{12}$ and $g_{21}$ to be 
independent.  
This implies
\begin{eqnarray}
\frac{\partial A}{\partial g_{11}}=\frac{g_{22}}{8A} && \frac{\partial A}{\partial g_{22}}=\frac{g_{11}}{8A} 
\label{eq:dAdgdiag}\\
\frac{\partial A}{\partial g_{12}}=-\frac{g_{21}}{8A} && \frac{\partial A}{\partial g_{21}}=-\frac{g_{12}}{8A}
\label{eq:dAdgoffdiag}
\end{eqnarray}
and therefore 
\begin{equation}
\frac{\partial A}{\partial L}=\frac{\partial A}{\partial g_{\mu\nu}}\frac{\partial g_{\mu\nu}}{\partial L}=
\frac{g_{12}}{4A}=\frac{L}{2}\cot\theta_{op}
\end{equation}
where we have used $g_{12}=l_1l_2\cos\theta_{op}$, 
$A=l_1l_2\sin\theta_{op}/2$, $g_{12}=(l_1^2+l_2^2-L^2)/2$, 
$g_{11}=l_1^2$ and $g_{22}=l_2^2$.
This is consistent with the standard Regge Calculus result for 
the field side of the Einstein field equations.  
Also, similar results apply for either of the other two edges 
if we change the metric so that its diagonal components do not have 
that edge.  If we do not change the metric, we find, for e.g., for $l_1$, 
\begin{equation}
\frac{\partial A}{\partial l_1}=\frac{1}{4}(l_2\csc\theta_{op}-l_1\cot\theta_{op})
\label{eq:dlalt}
\end{equation}
where $\theta_{op}$ is still the angle opposite edge $L$.
Also, one can ``go the other way'', and calculate $\frac{\partial A}{\partial g_{12}}$ via
\begin{equation}
\frac{\partial A}{\partial g_{12}}=\frac{\partial A}{\partial L}\frac{\partial L}{\partial L^2}\frac{\partial L^2}{\partial g_{12}}=
\left(\frac{L}{2}\cot\theta_{op}\right)\left(\frac{1}{2L}\right) (-1)=
-\frac{\cot\theta_{op}}{4}
\end{equation}
which is consistent with Eq.~(\ref{eq:dAdgoffdiag}), 
where we have used $L^2=g_{11}+g_{22}-g_{12}-g_{21}$ as well as 
$l_1^2=g_{11}$ and $l_2^2=g_{22}$.
So, for each edge corresponding to a $G^{\gamma}_{\;\;\gamma^\prime}$, 
it is easiest to vary each of the attached hinges with respect to 
the metric none of whose diagonal elements corresponds to that edge.

Next we show that $G^{\gamma}_{\;\;\gamma^\prime}$ is proportional to 
$A \, \delta$ for $\gamma =\gamma^\prime$, and zero otherwise, independently
of which edges were chosen to form the diagonal elements of the metric.  
So, the contribution from each hinge to the Einstein tensor for an 
edge is independent of the form of the metric \textit{if} one is 
considering the one index up and one index down form of the Einstein 
tensor.  So, Regge Calculus, to be ``metric independent'', would 
choose this form of the Einstein tensor to work with.  In particular, 
this metric independence does not hold when considering $G_{\mu\nu}$ or 
$G^{\mu\nu}$.  With a metric independent form one can have some sense of 
justification when one adds the various hinge contributions to edge L's 
$G^L_L$; $G^{LL}$ and $G_{LL}$, not having metric independence, are 
undefined, or, at least, non-unique.  Interestingly, the variation of 
$L$ contributes only to $G^{l_1l_1}$ and $G^{l_2l_2}$, but not 
$G^{LL}$; $G^{LL}$ receives contributions from varying $l_1$ and $l_2$.  
In terms of the number of $G^\mu_{\;\;\nu}$'s, we are consistent:  there 
would be three $g_{\rho\sigma}$'s per triangle, giving three 
$G^\mu_{\;\;\nu}$'s 
per triangle, which we take to be 
$G^{l_1}_{\;\; l_1}$, $G^{l_2}_{\;\; l_2}$ and $G^L_{\;\; L}$.

A crucial question at this point is whether the form of the Einstein 
tensor, $L\cot\theta_{op}/2$, is metric independent.  This form is 
obtained by varying with respect to edge length $L$, which is similar 
to varying with respect to edge length $L^2$ or $l_1\cdot l_2$.  Hence, 
this form most closely corresponds to either $G^{LL}$ or $G^{12}$, and 
therefore is not metric independent.  This can also be seen by 
noting our previous result of Eq.~({\ref{eq:dlalt}).

Next we calculate $G^{l_1}_{\;\; l_1}$ and obtain
\begin{equation}
G^{l_1}_{\;\; l_1}=g_{11}G^{11}+g_{12}G^{12}
=
\left[l_1^2\left(\frac{g_{22}}{8A}\right)+
(l_1l_2\cos\theta_{op})\left(\frac{-l_1l_2\cos\theta_{op}}{8A}\right)\right]
2\delta
              =A\delta
\end{equation}
which shows that our procedure is consistent.

\vskip 30pt
\newsection{General Form of the Bianchi Identity}
\hspace*{\parindent}

In the previous sections on three-dimensional (Section 4)
and four-dimensional (Section 5) rotations
an exact and explicit form of the lattice Bianchi identity was given
in terms of product of rotations along homotopically trivial
paths.
Later on in the paper an expression for the Riemann tensor and its dual
were proposed, which among other properties,
correctly reproduce the Regge field equations.
In this section we will show that if one proceeds from the continuum
Bianchi identities and uses the above given expression for
the lattice Riemann tensor then the following is true.
Firstly, the resulting lattice Bianchi identities coincide with their
weak field counterparts discussed in the sections on 3-d and
4-d rotations.
Secondly, when the infinitesimal form of the rotation matrix is
promoted to the correct finite rotation expression (of Sections 4 and 5),
the same exact identities derived previously are obtained.
These result presented here are therefore intended to bring out one more time
the close relationship between the Riemann tensor (in terms of
which the {\it continuum} identities are naturally formulated) and the
finite rotation matrices (in terms of which the exact {\it lattice}
identities are formulated). A lattice expression for the Riemann tensor
provides a natural bridge between these two different realms.

We will start here by considering the lattice Einstein tensor.
Since the Einstein tensor's components live on edges, derivatives of the 
Einstein tensor must occur at vertices.  Since the Einstein tensor is 
discontinuous at vertices, the easiest  procedure is to integrate the 
Bianchi identities over a d-volume around a vertex and divide by that 
d-volume, at least in the case of small deficit angles.\footnote{The 
assumption of a flat coordinate system, so that edges 
have definite directions and so that we will be able to use Gauss' theorem, 
introduces errors of ${\mathcal{O}}(\delta^2)$.}  
Then, the result is generalized to arbitrary deficit angles.
So, we begin by deriving the un-contracted, the partially 
contracted, and the completely contracted Bianchi identities in 
terms of the {\it dual} Riemann tensor, which will be the dual of the 
local Riemann curvature tensor as opposed to the global form, for 
reasons discussed earlier.
In the continuum the un-contracted Bianchi identities read

\begin{equation}
R^{\alpha^\prime\beta^\prime}_{\;\;\;\;\;\;\alpha\beta ;\gamma}+
R^{\alpha^\prime\beta^\prime}_{\;\;\;\;\;\;\beta\gamma ;\alpha}+
R^{\alpha^\prime\beta^\prime}_{\;\;\;\;\;\;\gamma\alpha ;\beta} \; = \; 0
\end{equation}
The partially contracted and completely contracted Bianchi identities 
are obtained via contraction of $\alpha$ with $\alpha^\prime$ or/and 
$\beta$ with $\beta^\prime$.  
Using the d-dimensional dual Riemann tensor, one can write the
un-contracted Bianchi identities into a divergence form via
\begin{equation}
\frac{1}{(d-2)!}
[(\epsilon_{\alpha\beta\mu_1...\mu_{d-2}}\tilde{R}^{\alpha^\prime\beta^\prime
                                                  \mu_1...\mu_{d-2}})_{;\gamma}
+(\epsilon_{\beta\gamma\nu_1...\nu_{d-2}}\tilde{R}^{\alpha^\prime\beta^\prime
                                                  \nu_1...\nu_{d-2}})_{;\alpha}
+(\epsilon_{\gamma\alpha\sigma_1...\sigma_{d-2}}
        \tilde{R}^{\alpha^\prime\beta^\prime\sigma_1...\sigma_{d-2}})_{;\beta}]
 \; = \; 0
\end{equation}
\begin{equation}
\frac{1}{(d-3)!}
[(\epsilon_{\alpha\beta\gamma\delta_1...\delta_{d-3}}
  \tilde{R}^{\alpha^\prime\beta^\prime\gamma\delta_1...\delta_{d-3}})_{;\gamma}
+(\epsilon_{\beta\gamma\alpha\delta_1...\delta_{d-3}}
  \tilde{R}^{\alpha^\prime\beta^\prime\alpha\delta_1...\delta_{d-3}})_{;\alpha}
+(\epsilon_{\gamma\alpha\beta\delta_1...\delta_{d-3}}
  \tilde{R}^{\alpha^\prime\beta^\prime\beta\delta_1...\delta_{d-3}})_{;\beta}]
 \; = \; 0
\end{equation}
\begin{equation}
\frac{1}{(d-3)!}
(\epsilon_{\alpha\beta\gamma\delta_1...\delta_{d-3}}
\tilde{R}^{\alpha^\prime\beta^\prime\lambda\delta_1...\delta_{d-3}})_{;\lambda}
 \; = \; 0
\end{equation}
Let us first consider the case of small deficit angles.
The next step is (1) to integrate over and divide by $^dV(v)$, the d-volume surrounding vertex 
$v$, (2) to use Gauss' theorem, and (3) break up the (d-1)-dimensional surface integral into
(d-1)-dimensional surfaces composed of (a) the 2-d Voronoi rotation areas
and (b) the 
(d-3)-dimensional portions of the hinges' surface volumes
(the $^{d-3}V_h$'s).
One finds
\begin{equation}
\frac{1}{(d-3)! \; {}^dV(v)}
\sum_{h\supset v}
\int_{^{d-1}V_h}
\!\!\!\!\!\!
\; \epsilon_{\alpha\beta\gamma\delta_1...\delta_{d-3}}
\; \tilde{R}^{\alpha^\prime\beta^\prime\lambda\delta_1...\delta_{d-3}} \; 
\hat{n}_{\lambda}\; \sqrt{|g|} \; d^{d-1}x \; = \; 0
\label{eq:bin}
\end{equation}
Rewriting $\tilde{R}$ using Eq.~(\ref{eq:riem}) and a generalized form 
of Eq.~(\ref{eq:dual}), letting the index $\lambda\rightarrow\lambda_0$, and 
defining the volume tensor
\begin{equation}
V_h^{\rho_0\rho_1...\rho_{d-3}}\equiv
         \frac{1}{(d-2)!}\sum_{(i_0,i_1,...,i_{d-3})={\mathcal{P}}
                               (0,1,...,d-3)}(-1)^{\mathcal{P}}
\; l_{h,i_0}^{\rho_0}l_{h,i_1}^{\rho_1}...l_{h,i_{d-3}}^{\rho_{d-3}}
\label{eq:Vh}
\end{equation}
where $(-1)^{\mathcal{P}}$ is the sign of the permutation, we find
the integrand to be
\begin{equation}
\epsilon_{\alpha\beta\gamma\delta_1...\delta_{d-3}}
\left[
\frac{\delta_h \; U^{\alpha^\prime\beta^\prime} 
      {}^{d-2}V_h^{\lambda_0\delta_1...\delta_{d-3}}}
     {A_\Gamma \; {}^{d-2}V_h}
\right]
\left[
\frac{\epsilon_{\mu\nu\lambda_0\lambda_1...\lambda_{d-3}}A^{\mu\nu}_\Gamma 
(l_{h,1}-l_{h,0})^{\lambda_1}...(l_{h,d-3}-l_{h,0})^{\lambda_{d-3}}}
     {A_\Gamma \; {}^{d-3}V_h}
\right]
\end{equation}
Note that for each hinge the above expression is independent of the choice of 
assignment of the labels $l_{h,0},l_{h,1},...,l_{h,d-3}$ to particular 
edges of the hinge.
Furthermore when Eq.~(\ref{eq:Vh}) is substituted into the 
above equation, each term in Eq.~(\ref{eq:Vh})'s sum can be replaced 
with $+l_0^{\lambda_0}l_1^{\delta_1}...l_{d-3}^{\delta_{d-3}}$, so that 
the contraction of edges with the 2-nd Levi-Civita tensor always gives 
$\pm A_\Gamma {}^{d-2}V_h$ 
where (4) we define $A^{\alpha\beta}_\Gamma$ such that 
the minus sign is always chosen 
(so as to be consistent with counter-clockwise rotations in 3-d).
After canceling the integration with $A_\Gamma {}^{d-3}V_h$, Eq.~(\ref{eq:bin}) 
can then be rewritten as 
\begin{equation}
-\frac{\epsilon_{\alpha\beta\gamma\delta_1...\delta_{d-3}}}{(d-3)! \; {}^dV(v)}
\sum_{h\supset v}
\delta_h \; U^{\alpha^\prime\beta^\prime}
\sum_{h_b\subset h}
l_{h,1}^{\delta_1} \; ... \; l_{h,d-3}^{\delta_{d-3}} \; = \; 0
\end{equation}
where $h_b$ refers to a ``hinge base'', and is a (d-3)-simplex, and hence 
necessarily a ``base'' to the hinges containing it.  
One can then invert the sums to find
\begin{equation}
-\frac{\epsilon_{\alpha\beta\gamma\delta_1...\delta_{d-3}}}{(d-3)! \; {}^dV(v)}
\sum_{h_b\supset v}
(\sum_{h\supset h_b}
\delta_h \; U_{h,h_b}^{\alpha^\prime\beta^\prime})
\; l_{h_b,1}^{\delta_1}...l_{h_b,d-3}^{\delta_{d-3}} \; = \; 0
\end{equation}
where a unique labeling assignment has been chosen for each hinge base, 
Here $U_{h,h_b}^{\alpha^\prime\beta^\prime} =\pm U^{\alpha^\prime\beta^\prime}$, 
the plus or minus sign chosen when $\{ l_{hb,1},...,l_{h_b,d-3}\}$ is, 
respectively, an even or odd permutation of $\{ l_{h,1},...,l_{h,d-3}\}$.  
I.e., the sign of $U_{h,h_b}^{\alpha^\prime\beta^\prime}$ is chosen such 
that 
$\epsilon_{\alpha\beta\gamma\delta_1...\delta_{d-3}}U_{h,h_b}^{\alpha\beta}
l_0^\gamma l_{h_b,1}^{\delta_1}...l_{h_b,d-3}^{\delta_{d-3}}<0$ for one, 
and hence for all via projection to 3-d, $h\supset h_b$.
\footnote{Here,
we replace $A^{\alpha^\prime\beta^\prime}_\Gamma$ with 
$A^{\alpha^\prime\beta^\prime}_{\Gamma ,h_b}\propto 
+U_{h,h_b}^{\alpha^\prime\beta^\prime}$.}

Using as the equation for $V_{h_b}^{\delta_1...\delta_{d-3}}$ 
a (d-3)-dimensional version of Eq.~(\ref{eq:Vh}), the above equation 
can then be rewritten as
\begin{equation}
-\frac{\epsilon_{\alpha\beta\gamma\delta_1...\delta_{d-3}}}{^dV(v)}
\sum_{h_b\supset v}
(\sum_{h\supset h_b}
\delta_h \; U_{h,h_b}^{\alpha^\prime\beta^\prime}) \;
V_{h_b}^{\delta_1...\delta_{d-3}} \; = \; 0
\label{eq:ucbismall}
\end{equation}
which for the special cases of three and four dimensions
reduces to the approximate small deficit angle expressions
discussed previously in the sections on 3d (Section 4)
and 4d (Section 5) rotations. 
Using 
$\sum\delta_h \; U^{\alpha^\prime\beta^\prime} \approx 
\sum (Rot -1)^{\alpha^\prime\beta^\prime} \approx 
\{\prod [1+(Rot-1)]-1\}^{\alpha^\prime\beta^\prime} = 
[\prod (Rot)-1]^{\alpha^\prime\beta^\prime}$, 
where ``$\approx$'' symbolizes equalities to first order in deficit angles, 
the above equation generalizes to, for arbitrary deficit 
angles,
\footnote{Here, $\prod (Rot)$ is a product of mixed tenors with 
the first index up and the second down; however the first index of the 
last rotation matrix (on the left), $\alpha^\prime$, 
and the 2nd index of the first rotation matrix, $\beta^\prime$, can 
be either up or down.} 
\begin{equation}
-\, \frac{\epsilon_{\alpha\beta\gamma\delta_1...\delta_{d-3}}}{^dV(v)}
\sum_{h_b\supset v} \;
[\prod_{h\supset h_b}(Rot_{h,h_b})-1]^{\alpha^\prime\beta^\prime}
\; V_{h_b}^{\delta_1...\delta_{d-3}} \; = \; 0
\label{eq:ucbirot}
\end{equation} 
where the product of rotations is over a null path, and where 
$\beta^\prime$ and $\alpha^\prime$ refer to, respectively, the first 
index of the first rotation matrix and the last index of the last 
rotation matrix.
In the special case of three and four dimensions
the above result is equivalent to the exact expressions
discussed previously in the sections on 3d (Section 4) and 4d (Section 5)
rotations. 
\footnote{In the Lorentzian case, the above equations have a 
plus sign; each term of the above equation gets an additional minus sign 
either because 
of the reason noted in the footnote to Eq.~(\ref{eq:riem}) or because 
the presence of the time component in a hinge would otherwise cause 
a reversal of the direction (and sign) of 
$A_{\Gamma ,h_b}^{\alpha^\prime\beta^\prime}$ 
due to the change in sign of 
$\epsilon_{\alpha\beta\gamma\delta_1...\delta_{d-3}}U_{h,h_b}^{\alpha\beta}
V_h^{\gamma\delta_1...\delta_{d-3}}$.}

Let us add a few comments regarding the result just obtained.
The coordinate system in which this equation is written 
is one of the possible global coordinate systems described in the
section on Geometric Setups (Section 3).
By taking the Levi-Civita 
tensor out of the $\sum_{h_b\supset v}$, we are taking the Levi-Civita 
tensor to be defined at vertices such as $v$.\footnote{We needn't have 
done this; however, taking it out of $\sum_{h\supset h_b}$ or 
$\prod_{h\supset h_b}$ avoids unnecessary complications.}  Now, 
one can ask how we can use a d-dimensional Levi-Civita tensor when 
in our derivation of ${\mathcal{G}}_{l}^{\;\; l}$ we only used 
$(d-2)\times (d-2)$ metrics; the answer is that there we were 
considering Voronoi volumes which had two dimensions not set by edge 
lengths, whereas here all dimensions of every volume emanating from 
vertex $v$ involve edge lengths.

Also, the partially and fully contracted 
Bianchi identities, now for arbitrary deficit angles, are easily obtained 
via contractions.
The above Bianchi identities can be 
converted to scalar form via dot products with $d$ d-vectors which form 
a d-volume in our going around coordinate system discussed in Section 3.  
Do note that, when one changes going around coordinate systems, dot 
products do change,
so that even here, what is a Bianchi identity for one going 
around coordinate system may not be for another going around coordinate 
system.  Another (going around) coordinate system 
difference would be that $\Gamma$'s such as that in 
Eq.~(\ref{eq:3dnullrotprior}) would be unity for some coordinate 
systems, and not for others, so that a different order of rotations, 
utilizing the $\Gamma$(s) which are unity, 
would be appropriate for these other coordinate systems.
Still, once the dot products are taken, the Bianchi identities are in 
terms of edge lengths, which are independent of coordinate system; hence, 
once dot products are taken, any Bianchi identity is true independent of 
the going around coordinate system.  
In summary, each going around coordinate system has its own 
Bianchi identities, which themselves can be projected into 
coordinate system independent equalities.  Of course, once one 
substitutes in for the deficit angles their explicit
expression in terms of edge lengths, one attains zero identically.
However, if one does not do this, these identities give relations 
between the deficit angles and squared edge lengths.

\vskip 30pt
\newsection{Bianchi Identities in Four Dimensions}
\hspace*{\parindent}

The discussion of the previous section was for a general dimension $d$.
In this section we will focus on the four-dimensional case, and
proceed from the continuum Bianchi identities to derive the
weak field expression for the lattice Bianchi identities. 
Once the resulting infinitesimal rotations are promoted to
finite lattice rotations, the resulting identities
coincide with the exact form given previously (Section 5).
At the end of this section we will then collect the explicit formulae for the
exact un-contracted, partially contracted, and fully
contracted lattice Bianchi identity, valid for arbitrary manifolds. 

In the four-dimensional case the un-contracted identity in the continuum reads
\begin{equation}
\frac{1}{2}
[(\epsilon_{\alpha\beta\mu\nu}
  \tilde{R}^{\alpha^\prime\beta^\prime\mu\nu})_{;\gamma}+
 (\epsilon_{\beta\gamma\mu\nu}
  \tilde{R}^{\alpha^\prime\beta^\prime\mu\nu})_{;\alpha}+
 (\epsilon_{\gamma\alpha\mu\nu}
  \tilde{R}^{\alpha^\prime\beta^\prime\mu\nu})_{;\beta}] \; = \; 0
\end{equation}
Now, defining the index $\delta$ such that $[\alpha\beta\gamma\delta ]\neq 0$ 
(for $d>4$, we choose \textit{one} set of indices, an \textit{ordered} set, 
$\vec{\delta}\equiv\delta_1\delta_2 ...\delta_{d-3}$, which satisfies 
$[\alpha\beta\gamma\vec{\delta}]\neq 0$), 
we find 
\begin{equation}
(\epsilon_{\alpha\beta\gamma\delta}
 \tilde{R}^{\alpha^\prime\beta^\prime\gamma\delta})_{;\gamma}+
(\epsilon_{\beta\gamma\alpha\delta}
 \tilde{R}^{\alpha^\prime\beta^\prime\alpha\delta})_{;\alpha}+
(\epsilon_{\gamma\alpha\beta\delta}
 \tilde{R}^{\alpha^\prime\beta^\prime\beta\delta})_{;\beta} \; = \; 0
\end{equation}
where, as we have stated, 
$\alpha$, $\beta$, $\gamma$ and $\delta$ are not summed anywhere in the 
equation.  
We can now rewrite the above equation as 
\begin{equation}
(\epsilon_{\alpha\beta\gamma\delta}
 \tilde{R}^{\alpha^\prime\beta^\prime\lambda\delta})_{;\lambda} \; = \; 0
\label{eq:theucbi}
\end{equation}
where $\lambda$ is summed over, and $\delta$ is not.  Actually, in 4-d, 
the above equation does not change even if $\delta$ is summed over all 
directions; for 
$d>4$, one simply multiplies by a factor of $1/(d-3)!$ to get the 
exact correspondence to the un-contracted Bianchi identities 
if one decides to sum on $\delta$ (which, for $d>4$, becomes 
$\vec{\delta}$, 
and the ``sum on $\delta$'' is the sum over all the $\delta$'s of 
$\vec{\delta}$, which has non-zero terms only when none of the 
$\delta$'s are equal to $\alpha$, $\beta$ or $\gamma$).

To get the partially contracted Bianchi identites, one simply either 
sets $\alpha =\alpha^\prime$ or $\beta =\beta^\prime$, and then sums 
over either $\alpha$ or $\beta$.
Setting 
$\beta =\beta^\prime$ gives
\begin{equation}
(\epsilon_{\alpha\beta\gamma\delta}
 \tilde{R}^{\alpha^\prime\beta\lambda\delta})_{;\lambda} \; = \; 0
\end{equation}
Here, we use the Einstein summation convention for all indices, 
so that for $d>4$, we would need a factor of $1/(d-3)!$ from 
Eq.~(\ref{eq:theucbi}).  In particular, we note that $\delta$ 
must be summed over, because we are summing over two values of 
$\beta$, and therefore $\delta$ must run over different indices 
(or, for $d>4$, different $\beta$'s implies that not all $\vec{\delta}$'s 
which give non-zero terms will be comprised of the same set of indices).
To get the fully contracted Bianchi identities, we set 
$\alpha =\alpha^\prime$ and sum over $\alpha$ 
in the partially contracted Bianchi identities to find 
\begin{equation}
(\epsilon_{\alpha\beta\gamma\delta}
 \tilde{R}^{\alpha\beta\lambda\delta})_{;\lambda} \; = \; 0
\end{equation}
where, once again, we would have the factor of $1/(d-3)!$ for $d>4$.

The next step is to interpret the un-contracted Bianchi identities, 
Eq.~(\ref{eq:theucbi}) in Regge Calculus.  Using the lattice expression
for the dual of the Riemann curvature tensor, Eq.~(\ref{eq:dualRReg}), 
we have

\begin{equation}
\frac{1}{6}
(\frac{\epsilon_{\alpha\beta\gamma\delta}}{^4V_h}
 \delta_h \; U^{\alpha^\prime\beta^\prime} A_h^{\lambda\delta})_{;\lambda} \; = \; 0
\end{equation}
We now integrate this over a vertex $v$, and divide by the sum of all 
Voronoi four-volumes surrounding that vertex, $^4V_v$.

\begin{equation}
\frac{1}{6}
\frac{1}{^4V_v}
\int_v
(\frac{\epsilon_{\alpha\beta\gamma\delta}}{^4V_h}
 \delta_h \; U^{\alpha^\prime\beta^\prime} A_h^{\lambda\delta})_{;\lambda}
\sqrt{|g|} \; d^4x
 \; = \; 0
\end{equation}
One can rewrite this in terms of edges, as the components of the 
Einstein tensor 
live on edges, and because the completely contracted Bianchi identities 
are written in terms of components of the Einstein tensor.

\begin{equation}
\frac{1}{6} \; 
\frac{1}{^4V_v}
\sum_{h\supset v}
\int_h
(\frac{\epsilon_{\alpha\beta\gamma\delta}}{^4V_h}
\; \delta_h \; U^{\alpha^\prime\beta^\prime} A_h^{\lambda\delta})_{;\lambda} \;
\sqrt{|g|} \; d^4x
 \; = \; 0
\label{eq:bisumhint}
\end{equation}
Since each hinge appears on two edges, we find

\begin{equation}
\frac{1}{2 \cdot 6}
\frac{1}{^4V_v}
\sum_{l\supset v}
\sum_{h\supset l}
\int_h
(\frac{\epsilon_{\alpha\beta\gamma\delta}}{^4V_h}
\; \delta_h \; U^{\alpha^\prime\beta^\prime} A_h^{\lambda\delta})_{;\lambda} \;
\sqrt{|g|} \; d^4x
 \; = \; 0
\label{eq:bisumlhint}
\end{equation}
which implies that 
the dual Riemann tensor for an edge is 
one half of 
the sum of the dual Riemann tensor for all hinges containing that edge.  
Using Gauss' theorem, one then finds
\begin{equation}
\frac{1}{6}
\frac{1}{^4V_v}
\sum_{l\supset v}
\sum_{h\supset l}
\int_h
(\frac{\epsilon_{\alpha\beta\gamma\delta}}{^4V_h}
\; \delta_h \; U^{\alpha^\prime\beta^\prime} A_h^{\lambda\delta}) \; \hat{n}_{\lambda}
\sqrt{|g|} \; d^3x
 \; = \; 0
\end{equation}
Noting that the radial direction is contained within the hinge, 
one can integrate over the loop area, leaving the only integration 
variable to be the direction $\pm (L-l)$, where $L$ is the other edge 
of the hinge which also contains (and is directed outward from) the 
vertex $v$.
$L-l$ here is the unique direction, 
up to a sign, perpendicular to the surface normal without any component
contained within the loop area.  We choose $+(L-l)$ to ensure consistency 
throughout the sums.  One has therefore
\begin{equation}
\frac{1}{6}
\frac{1}{^4V_v}
\sum_{l\supset v}
\sum_{h\supset l}
\int_h
(A_\Gamma\frac{\epsilon_{\alpha\beta\gamma\delta}}{^4V_h}
\; \delta_h \; U^{\alpha^\prime\beta^\prime} A_h^{\lambda\delta}) \; \hat{n}_{\lambda}
\sqrt{|g|} \; ds
 \; = \; 0
\end{equation}
Noting that $A_h^{\lambda\delta}=L^\lambda l^\delta -
l^\lambda L^\delta$ (where we have chosen a sign convention for $A_h$), 
one can write the dot product $L\cdot\hat{n}$ as

\begin{equation}
\hat{n}\cdot L=\frac
{[\epsilon_{\mu\nu\lambda\sigma}A_\Gamma^{\mu\nu}(l-L)^\sigma ]L^\lambda}
{A_\Gamma |L-l|}
=\frac
{\epsilon_{\mu\nu\lambda\sigma}A_\Gamma^{\mu\nu}l^\sigma L^\lambda}
{A_\Gamma |L-l|}
=\frac{\pm 2A_h}{|L-l|}
\end{equation}
Then using this result and a similar one for $l\cdot\hat{n}$ one obtains

\begin{equation}
\frac{1}{6}
\frac{1}{^4V_v}
\sum_{l\supset v}
\sum_{h\supset l}
\frac{\epsilon_{\alpha\beta\gamma\delta}}{^4V_h}
\; \delta_h \; U^{\alpha^\prime\beta^\prime} A_\Gamma A_h \; (l-L)^\delta
 \; = \; 0
\end{equation}
which is equivalent to
\begin{equation}
\frac{1}{^4V_v}
\sum_{l\supset v}
\epsilon_{\alpha\beta\gamma\delta}
(\pm )
\sum_{h\supset l}
 \delta_h \; U^{\alpha^\prime\beta^\prime} (l-L)^\delta
 \; = \; 0
\end{equation}
where the $\pm$ sign will indicate whether the hinges, for each particular 
hinge base, are gone around in a counter-clockwise or clockwise fashion.  
This is then further simplified when one considers that the contribution 
for each edge is the same whether it be in its $L$ (non-hinge base) 
contribution, or its $l$ (hinge base) contribution, noting that 
$l\leftrightarrow L$ leads to $\pm \rightarrow \mp$.  Indeed, given that 
we started with a sum over hinges, Eq.~(\ref{eq:bisumhint}), 
and broke that down into a sum over edges, Eq.~(\ref{eq:bisumlhint}), 
each edge's contribution to the hinge must be identical.  So, we can take 
each edge's $l$ contribution, double it, and omit its $L$ contribution to 
find for small deficit angle
\begin{equation}
\frac{1}{^4V_v}
\sum_{l\supset v}
\epsilon_{\alpha\beta\gamma\delta}
(\pm )
\sum_{h\supset l}
 \; \delta_h \; U^{\alpha^\prime\beta^\prime} l^\delta
 \; = \; 0
\end{equation}
The partially contracted and completely contracted forms 
are then easy to obtain.
The partially contracted form can be written for small deficit angle
as either 
\begin{equation}
\frac{1}{^4V_v}
\sum_{l\supset v}
\sum_{h\supset l}
\epsilon_{\alpha\beta\gamma\delta}
 \; \delta_h \; U^{\alpha^\prime\beta} (L-l)^\delta
 \; = \; 0
\end{equation}
or
\begin{equation}
\frac{1}{^4V_v}
\sum_{l\supset v}
\sum_{h\supset l}
\epsilon_{\alpha\beta\gamma\delta}
 \; \delta_h \; U^{\alpha\beta^\prime} (L-l)^\delta
 \; = \; 0
\end{equation}
and the fully contracted form for small deficit angle is
\begin{equation}
\frac{1}{^4V_v}
\sum_{l\supset v}
\sum_{h\supset l}
\epsilon_{\alpha\beta\gamma\delta}
 \; \delta_h \; U^{\alpha\beta} (L-l)^\delta
 \; = \; 0
\end{equation}
We note that each fixed $l$ term in $\sum_l$ can be taken to 
correspond to one term in (what was) the divergence, because each 
edge length corresponds to one component of the Einstein tensor 
$G^{\lambda}_{\;\;\gamma}$.

Next we note that these small deficit angle forms can all
be rewritten in terms of rotation 
matrices.  For example, the un-contracted form can be written 
as
\begin{equation}
\frac{1}{^4V_v}
\sum_{l\supset v}
\sum_{h\supset l}
\epsilon_{\alpha\beta\gamma\delta}
 (Rot_h -1)^{\alpha^\prime\beta^\prime} (L-l)^\delta
 \; = \; 0
\end{equation}
and for arbitrary deficit angles, this generalizes to
\begin{equation}
\frac{1}{^4V_v}
\sum_{l\supset v}
\epsilon_{\alpha\beta\gamma\delta}
 [\prod_{h\supset l}(Rot_h)-1]^{\alpha^\prime\beta^\prime} (L-l)^\delta
 \; = \; 0
\label{eq:ucbi4d}
\end{equation}
where $\beta^\prime$ refers to the second index of the first rotation 
matrix, and $\alpha^\prime$ refers to the first index of the last 
rotation matrix.
The above expression coincides with the exact result obtained previously
in the section on finite rotations in four dimensions (Section 5).
The partially contracted Bianchi identities can then 
be written for arbitrary deficit angles as either 
\begin{equation}
\frac{1}{^4V_v}
\sum_{l\supset v}
\epsilon_{\alpha\beta\gamma\delta}
 [\prod_{h\supset l}(Rot_h)-1]^{\alpha^\prime\beta} (L-l)^\delta
 \; = \; 0
\label{eq:pcbi4d}
\end{equation}
or
\begin{equation}
\frac{1}{^4V_v}
\sum_{l\supset v}
\epsilon_{\alpha\beta\gamma\delta}
 [\prod_{h\supset l}(Rot_h)-1]^{\alpha\beta^\prime} (L-l)^\delta
 \; = \; 0
\end{equation}
depending on the choice of indices.
Finally the completely contracted Bianchi identities can be written for
arbitrary deficit angle as 
\begin{equation}
\frac{1}{^4V_v}
\sum_{l\supset v}
\epsilon_{\alpha\beta\gamma\delta}
 [\prod_{h\supset l}(Rot_h)-1]^{\alpha\beta} (L-l)^\delta
 \; = \; 0
\label{eq:fcbi4d}
\end{equation}
We can see that this has the correct units, $1/( {}^3V )$, by taking the 
metric, and hence the Levi=Civita tensor, to be dimensionless.

\vskip 30pt
\newsection{Bianchi Identities in Three Dimensions}
\hspace*{\parindent}

Let us repeat the above construction in three dimensions, without
going through the details which very much parallel what was
done in the four dimensional case just discussed.
For $d=3$, since the hinges are the edges, we get for the completely 
un-contracted Bianchi identities, valid for arbitrary deficit angles,
\begin{equation}
\frac{1}{^3V_v}
\; \epsilon_{\alpha\beta\gamma}
 \; [\prod_l(Rot_l)-1]^{\alpha^\prime\beta^\prime}
 \; = \; 0
\label{eq:ucbi3d}
\end{equation}
Then the 3-d partially contracted Bianchi identites 
for arbitrary deficit angles can be written as either 
\begin{equation}
\frac{1}{^3V_v}
\; \epsilon_{\alpha\beta\gamma}
 \; [\prod_l(Rot_l)-1]^{\alpha^\prime\beta}
 \; = \; 0
\label{eq:pcbi3d}
\end{equation}
or
\begin{equation}
\frac{1}{^3V_v}
\; \epsilon_{\alpha\beta\gamma}
 \; [\prod_l(Rot_l)-1]^{\alpha\beta^\prime}
 \; = \; 0
\end{equation}
and finally the 3-d completely contracted Bianchi identities
for arbitrary deficit angles read
\begin{equation}
\frac{1}{^3V_v}
\; \epsilon_{\alpha\beta\gamma}
 \; [\prod_l(Rot_l)-1]^{\alpha\beta}
 \; = \; 0
\label{eq:fcbi3d}
\end{equation}
These expressions coincide with the exact result obtained previously in the
section on finite rotations in three dimensions (Section 4).
In three dimensions and for small deficit angles these results reduce to 
\begin{equation}
\frac{1}{^3V_v}
\sum_{l\supset v}
\epsilon_{\alpha\beta\gamma}
 \; \delta_l \; U^{\alpha^\prime\beta^\prime}
 \; = \; 0
\end{equation}
The 3-d partially contracted Bianchi identites can be then be
written, again for small deficit angles, as either 
\begin{equation}
\frac{1}{^3V_v}
\sum_{l\supset v}
\epsilon_{\alpha\beta\gamma}
 \; \delta_l \; U^{\alpha^\prime\beta}
 \; = \; 0
\end{equation}
or
\begin{equation}
\frac{1}{^3V_v}
\sum_{l\supset v}
\epsilon_{\alpha\beta\gamma}
 \; \delta_l \; U^{\alpha\beta^\prime}
 \; = \; 0
\end{equation}
depending on the choice of indices.
Finally for small deficit angles
the 3-d completely contracted Bianchi identities are
\begin{equation}
\frac{1}{^3V_v}
\; \epsilon_{\alpha\beta\gamma}
\sum_{l\supset v}
 \; \delta_l \; U^{\alpha\beta}
 \; = \; 0
\end{equation}

\vskip 30pt
\newsection{Bianchi Identities in Dimensions Greater than Four}
\hspace*{\parindent}

For dimensions $d>4$, our $4-d$ results are generalized after recalling 
that, in the original integration, each hinge appears once, 
so that each hinge should appear once in our results as well.  
For each hinge, all of its $(d-2)$ edges bordering vertex $v$ appear in 
$\sum_l$ once, so we need 
to divide by $(d-2)$.  Also, for a fixed edge contained within a hinge, there 
are 
$\left(\begin{array}{c}d-3\\d-4\end{array}\right)=d-3$ 
hinge bases ((d-3)-dimensional simplexes which are labeled by $h_b$), 
which are summed over which that hinge contains.  
So, we must also divide by $(d-3)$.
For convenience, the order of $L_1,L_2,...,L_{d-3}$ can 
be chosen such that 
$\epsilon_{\alpha\beta\gamma\vec{\delta}}
\, Rot_h^{\alpha\beta}l^\gamma \, (L-l)^{\vec{\delta}} > 0$.
Therefore for $d>4$ the un-contracted Bianchi identities for small deficit angles
read
\begin{equation}
\frac{1}{(d-2)(d-3)}
\frac{1}{^dV_v}
\sum_{l\supset v}
\sum_{h_b\supset l}
\epsilon_{\alpha\beta\gamma\vec{\delta}} \;
 [\sum_{h=(h_b,L)} \delta_h \; U^{\alpha\beta} 
(l_1-l)^{\delta_1}(l_2-l)^{\delta_2}...(l_{d-4}-l)^{\delta_{d-4}}
(l-L)^{\delta_{d-3}}
 \; = \; 0
\label{eq:dgt4bia}
\end{equation}
where the last sum can be viewed either as over $h\supset h_b$ or over 
edges $L$ which, along with $h_b$, form a hinge.  Taking the latter view, 
we can note that, for a fixed term in this triple sum, we should get the 
same term when $l\leftrightarrow L$. This, indeed, is the reason we 
were able to divide by $(d-2)$ in Eq.~(\ref{eq:dgt4bia}).
So one can simply double the above result and let 
$(l-L)^{\delta_{d-3}}\rightarrow l^\delta_{d-3}$, which dramatically 
simplifies the answer to
\begin{equation}
\frac{2}{(d-2)(d-3)}
\frac{1}{^dV_v}
\sum_{l\supset v}
\sum_{h_b\supset l}
\epsilon_{\alpha\beta\gamma\vec{\delta}} \;
 [\sum_{h=(h_b,L)} \delta_h \; U^{\alpha\beta} 
l_1^{\delta_1}l_2^{\delta_2}...l_{d-4}^{\delta_{d-4}}
l^{\delta_{d-3}}
 \; = \; 0
\end{equation}
Now, as opposed to $\frac{1}{(d-2)(d-3)}\sum_{l\supset v}\sum_{h_b\supset l}$ 
we could have written 
just $\frac{1}{d-2}\sum_{h_b\supset v}$, 
but this would not explicitly 
involve the edge lengths.  And, a sum over edge lengths is most directly 
related to a divergence because, as mentioned earlier, 
each fixed $l$ term in $\sum_l$ can be taken to 
correspond to one term in (what was) the divergence, because each 
edge length corresponds to one component of the Einstein tensor. 
Additionally, the identities 
would not look as neat, as there would not be a natural edge to subtract 
from the other hinge edges to get a hinge's $(d-3)$-surface volume.

The above formulas were for small deficit angles.
For arbitrary deficit angles, we find the un-contracted 
Bianchi identities to be

\begin{equation}
\frac{1}{(d-2)(d-3)}
\frac{1}{^dV_v}
\sum_{l\supset v}
\sum_{h_b\supset l}
\epsilon_{\alpha\beta\gamma\vec{\delta}}
\; [(\prod_{h\supset h_b}Rot_h) -1]^{\alpha^\prime\beta^\prime} 
(L-l)_{h_b,v}^{\vec{\delta}}
 \; = \; 0
\label{eq:ucbigt4d}
\end{equation}
The partially contracted Bianchi identities can be written as either 
\begin{equation}
\frac{1}{(d-2)(d-3)}
\frac{1}{^dV_v}
\sum_{l\supset v}
\sum_{h_b\supset l}
\epsilon_{\alpha\beta\gamma\vec{\delta}}
\; [(\prod_{h\supset h_b}Rot_h) -1]^{\alpha^\prime\beta} 
(L-l)_{h_b,v}^{\vec{\delta}}
 \; = \; 0
\label{eq:pcbigt4d}
\end{equation}
or
\begin{equation}
\frac{1}{(d-2)(d-3)}
\frac{1}{^dV_v}
\sum_{l\supset v}
\sum_{h_b\supset l}
\epsilon_{\alpha\beta\gamma\vec{\delta}}
\; [(\prod_{h\supset h_b}Rot_h) -1]^{\alpha\beta^\prime} 
(L-l)_{h_b,v}^{\vec{\delta}}
 \; = \; 0
\end{equation}
depending on the choice of indices,
and finally the completely contracted Bianchi identities can be written as 
\begin{equation}
\frac{1}{(d-2)(d-3)}
\frac{1}{^dV_v}
\sum_{l\supset v}
\sum_{h_b\supset l}
\epsilon_{\alpha\beta\gamma\vec{\delta}}
\; [(\prod_{h\supset h_b}Rot_h) -1]^{\alpha\beta} 
(L-l)_{h_b,v}^{\vec{\delta}}
 \; = \; 0
\label{eq:fcbigt4d}
\end{equation}

\vskip 30pt
\newsection{Conclusions}
\hspace*{\parindent}

In this paper we have derived an exact form for the Bianchi
identities in simplicial gravity.
In four dimensions these are given by 
Eqs.~(\ref{eq:ucbi4d}), (\ref{eq:pcbi4d}) and (\ref{eq:fcbi4d}) for
the un-contracted, partially contracted and fully contracted form
respectively.
In three dimensions the corresponding expressions are given by
Eqs.~(\ref{eq:ucbi3d}), (\ref{eq:pcbi3d}) and (\ref{eq:fcbi3d}), while
above four dimension the corresponding general results are
in Eqs.~(\ref{eq:ucbigt4d}), (\ref{eq:pcbigt4d}) and (\ref{eq:fcbigt4d}),
with an alternative but equivalent form of the un-contracted
Bianchi identity given in Eq.~(\ref{eq:ucbirot}). 
The explicit form of the rotation matrices appearing in the
above-quoted exact Bianchi identities was constructed explicitly
and presented in Sections 4 (three dimensional case) and 5 (four
dimensional case). 

While fairly unwieldy in their explicit form, these
identities can be shown to reduce to their known weak field
expression in the limit of small curvatures. 
They provide an explicit, local relationship between deficit 
angles belonging to neighboring simplices.
Their existence can be viewed as a consequence of the local invariance
of the Regge action under small gauge deformations of
edge lengths emanating from a vertex, just as the
continuum Bianchi identity can be derived from the local
gauge invariance of the gravitational action.

The relationship between the lattice Bianchi identities
and the Regge lattice equations of motion has been investigated as well.
In the continuum the contracted Bianchi identities ensure the consistency
of the gravitational field
equations. One would expect that the same should be true
on the lattice, in the sense that a lattice ``covariant divergence''
of the lattice field equations would identically yield zero,
as a consequence of the lattice Bianchi identity.
We have shown that that is indeed the case in the lattice theory.

When in the Regge lattice case we go from the discrete Einstein-Hilbert
action and its equations of motion
to the fully contracted Bianchi identities, 
and then write down an explicit form for the partially contracted
and un-contracted Bianchi identities as well,
we find that the relation between the three 
types of Bianchi identities in the Regge theory is basically
simply the contraction of indices, as in the continuum.   
Also, as the product of rotation matrices around 
a null path is critical in understanding the form of all three types of 
Bianchi identities, we have provided both a quick way to calculate
this product and an understanding of, in particular, its antisymmetric
components.
Finally by appropriately projecting the lattice Bianchi identities,
we have derived explicit expressions depending on edge
lengths squared only.

All of this is exact, valid for arbitrary deficit angles, though we have 
provided first order approximations to our results for the product of 
rotation matrices. 
An exact form for the lattice Bianchi identity should be useful
in a variety of context, including numerical schemes for classical and
quantum gravity. In the classical case, the accuracy
of four-dimensional time evolution codes could be checked by evaluating
the Bianchi identity along a time evolved trajectory.

\vspace{20pt}

\vspace{24pt}

{\bf Acknowledgements}

The authors wish to thank J. Hartle and R.M. Williams for very useful
discussions on the subject of this paper.

\vfill
\newpage
\end{document}